%% file: LUM-20-002_temp.tex
\pdfoutput=1
\documentclass[11pt,twoside,a4paper,cmspaper,final,collab]{cms-tdr}
\def\svnVersion{acb14f7-D}\def\svnDate{2025/03/05}\def\cmsCernNoTag{CERN-EP-2025-018}\def\cmsCernDate{\today}\def\cmsMessage{Submitted to the European Physical Journal C}
\begin{document}\cmsNoteHeader{LUM-20-002}

\newcommand{\PbPb}{\ensuremath{\mathrm{PbPb}}\xspace}
\newcommand{\sigmaVis}{\ensuremath{\sigma_{\text{vis}}}\xspace}

\newcommand{\Dx}{\ensuremath{\Delta x}\xspace}
\newcommand{\Dy}{\ensuremath{\Delta y}\xspace}
\newcommand{\Dxz}{\ensuremath{\Dx_0}\xspace}
\newcommand{\Dyz}{\ensuremath{\Dy_0}\xspace}
\newcommand{\Aeff}{\ensuremath{A_{\text{eff}}}\xspace}
\newcommand{\Sigx}{\ensuremath{\Sigma_x}\xspace}
\newcommand{\Sigy}{\ensuremath{\Sigma_y}\xspace}

\newcommand{\XY}{\ensuremath{x\text{--}y}\xspace}
\newcommand{\YX}{\ensuremath{y\text{--}x}\xspace}
\newcommand{\sigmab}{\ensuremath{\sigma_{\text{b}}}\xspace}
\newcommand{\usigmab}{\unit{\sigmab}}

\newcommand{\nbinvns}{\mbox{\ensuremath{\text{nb}^{-1}}}}
\newcommand{\chisq}{\ensuremath{\chi^2}\xspace}
\newcommand{\dof}{\ensuremath{\text{dof}}\xspace}
\newcommand{\chisqred}{\ensuremath{\chisq/\dof}\xspace}
\newcommand{\betast}{\ensuremath{\beta^\ast}\xspace}
\newcommand{\None}{\ensuremath{N_1}\xspace}
\newcommand{\Ntwo}{\ensuremath{N_2}\xspace}

\ifthenelse{\boolean{cms@external}}{
    \providecommand{\cmsTable}[1]{\resizebox{\linewidth}{!}{#1}}
    \providecommand{\cmsLeft}{upper\xspace}
    \providecommand{\cmsRight}{lower\xspace}
}{
    \providecommand{\cmsTable}[1]{#1}
    \providecommand{\cmsLeft}{left\xspace}
    \providecommand{\cmsRight}{right\xspace}
}
\newlength\cmsTabSkip\setlength\cmsTabSkip{0.5ex}

\cmsNoteHeader{LUM-20-002}
\title{Luminosity measurement for lead-lead collisions at \texorpdfstring{$\sqrtsNN=5.02\TeV$}{sqrt(sNN) = 5.02 TeV} in 2015 and 2018 at CMS}

\date{\today}

\abstract{Measurements of the luminosity delivered to the CMS experiment during the lead-lead data-taking periods in 2015 and 2018 are presented for the first time. The collisions were recorded at a nucleon-nucleon center-of-mass energy of 5.02\TeV; the 2018 data sample is three times larger than the 2015 data sample. Three subdetectors are used: the pixel luminosity telescope, the forward hadron calorimeters, and the fast beam conditions monitor. The absolute luminosity calibration is determined using the van der Meer technique that relies on transverse beam separation scans. The dominant sources of uncertainty are the transverse factorizability of the bunch density profiles and, in 2015, the difference between the results obtained using various detectors. The total uncertainty in the integrated luminosity, including the stability of the calibrated subdetector response over time, amounts to 3.0\% for 2015, 1.7\% for 2018, and 1.6\% for the combined data sample.}

\hypersetup{%
pdfauthor={CMS Collaboration},%
pdftitle={Luminosity measurement for lead-lead collisions at sqrt(sNN) = 5.02 TeV in 2015 and 2018 at CMS},
pdfsubject={CMS},%
pdfkeywords={CMS, PbPb, luminosity}}

\maketitle

\section{Introduction}

{\tolerance=2000
The measurement of the luminosity for lead-lead (\PbPb) collision data at a nucleon-nucleon center-of-mass energy $\sqrtsNN=5.02\TeV$ recorded at the CERN LHC interaction point (IP) 5 with the CMS detector in 2015 and 2018 is reported for the first time.
A good understanding of the luminosity is critical to reduce the systematic uncertainty in cross section measurements, \eg, Refs.~\cite{CMS:2024krd, CMS:2020aem}.
To determine the luminosity $\Lumi(t)$ at any given time $t$, we record the rate $R(t)$ of a specific subdetector (``luminometer'') suitable for the luminosity measurement, and use a prior knowledge of the visible cross section \sigmaVis of inelastic \PbPb scattering resulting in a count observed in that luminometer to obtain:%
\begin{equation}\label{eq:lumirel}
    \Lumi(t) = \frac{R(t)}{\sigmaVis}.
\end{equation}}
The \sigmaVis is related to the total inelastic cross section via a detector- and observable-dependent factor, and depends in particular on the fiducial (``visible'') phase space region of the luminometer.
A precise measurement of $\Lumi(t)$ then requires a precise calibration of \sigmaVis, as well as a monitoring of possible effects that bias the $R(t)$ measurement.

The absolute luminosity calibration, \ie, the measurement of \sigmaVis for all the luminometers employed in the luminosity measurement, is established with van der Meer (vdM) transverse beam separation ``scans''~\cite{vanderMeer:296752}.
By displacing (``scanning'') the two beams through one another in the plane transverse to the beam directions in a dedicated LHC fill, the effective width of the bunch-bunch collision area along the scanning direction is determined from rate measurements.
The analysis of pairs of such vdM scans performed in orthogonal directions together with other measured beam parameters facilitates a precise measurement of the luminosity per colliding bunch pair~\cite{Grafstrom:2015foa}.
The LHC experiments previously published luminosity calibration results with the vdM technique in proton-proton~\cite{ATLAS:2013tpw, Aaij:2014ida, ATLAS:2016fhk, CMS:2021xjt, ATLAS:2022hro}, proton-nucleus~\cite{Abelev:2014epa, ATLAS:2019ery}, and \PbPb~\cite{ATLAS:2020epq, ALICE:2022xir, ATLAS:2022ryk} collisions.

{\tolerance=800
Various luminometers are used at the CMS experiment to monitor and measure luminosity.
Redundant measurements whose comparisons can indicate detec\-tor-related disagreements, \eg, caused by electronic noise or radiation damage in sensors, greatly benefits the evaluation of the instrumentation response stability over long periods of time, and thus facilitates a precise estimation of the integrated luminosity.
\par}

The independent luminometers employed in these measurements in order of their importance
are: (i) the pixel luminosity telescope (PLT)~\cite{CMSBRIL:2022qin}, which is the primary source of luminosity data; (ii) the forward hadron calorimeter (HF)~\cite{CERN-LHCC-97-031}, which acts as a secondary source; and (iii) for 2018 the fast beam conditions monitor (BCM1F)~\cite{CMS:2016tsw, Hempel:2017nvn, Guthoff:2019cil}, which acts as tertiary source that is used if neither of the other two sources are available.
These luminometers are read out by an independent high-rate data acquisition (DAQ) system and serve as excellent real-time (``online'') monitors with a 1\% statistical accuracy in less than 1 second of data taking.
The stability of the luminometer responses is monitored during the entire \PbPb data-taking periods, which were in the months of November and December in both 2015 and 2018.
In 2018, the HF and PLT stabilities are additionally monitored using short vdM-like (``emittance'') scans that were performed regularly during the \PbPb collisions period.

The outline of this paper is as follows.
In Section~\ref{sec:cms}, a short description of the CMS detector and of the studied luminometers is provided.
The experimental setup to measure the luminosity is detailed in Section~\ref{sec:vdmScan}.
Section~\ref{sec:calibrations} describes various systematic effects that modify \sigmaVis, as well as the estimation of the corresponding uncertainties.
Including corrections for all systematic effects, the measured \sigmaVis values for the three luminometers are presented in Section~\ref{sec:sigvis}.
Additional uncertainties from cross-calibrating luminometers are evaluated in Section~\ref{sec:stability}.
The combined results of the luminosity measurement are given in Section~\ref{sec:syserr}.

\section{The CMS detector}
\label{sec:cms}

The CMS apparatus is a multipurpose, nearly hermetic detector, designed to study a broad range of phenomena in proton-proton and heavy ion collisions.
Its central feature is a superconducting solenoid of 6\unit{m} internal diameter, providing a magnetic field of 3.8\unit{T}.
Within the solenoid volume are a silicon pixel and strip tracker, a lead tungstate crystal electromagnetic calorimeter, and a brass and scintillator hadron calorimeter, each composed of a barrel and two endcap sections.
The HF calorimeters extend the pseudorapidity ($\eta$) coverage provided by the barrel and endcap detectors.
Muons are reconstructed in gas-ionization detectors embedded in the steel flux-return yoke outside the solenoid.
More detailed descriptions of the CMS detector, together with a definition of the coordinate system used and the relevant kinematic variables, are reported in Refs.~\cite{Chatrchyan:2008zzk, CMS:PRF-21-001}.

{\tolerance=2400
Events of interest are selected using a two-tiered trigger system.
The first level, composed of custom hardware processors, uses information from the calorimeters and
muon detectors to select events within a fixed latency of 4\mus~\cite{CMS:TRG-17-001}.
The second level, known as the high-level trigger, consists of a farm of processors running a version of the full event reconstruction software optimized for fast processing, and further reduces the event rate
before data storage~\cite{Khachatryan:2016bia, CMS:TRG-19-001}.
\par}

Primary-interaction vertices reconstructed with the CMS tracker system~\cite{tracking} are used to determine the center of the luminous region, required for the length scale calibration discussed in Section~\ref{sec:lsc}.
The corresponding data are recorded with the central CMS DAQ system, using a dedicated online filter~\cite{Khachatryan:2016bia} to select hadronic interactions.
This ``minimum bias'' selection with a bandwidth of about 12\unit{kHz} is based on coincidences between the trigger signals from the $\eta>0$ and $\eta<0$ HF calorimeters.

\begin{figure*}[!ht]
\centering
\includegraphics[width=\textwidth]{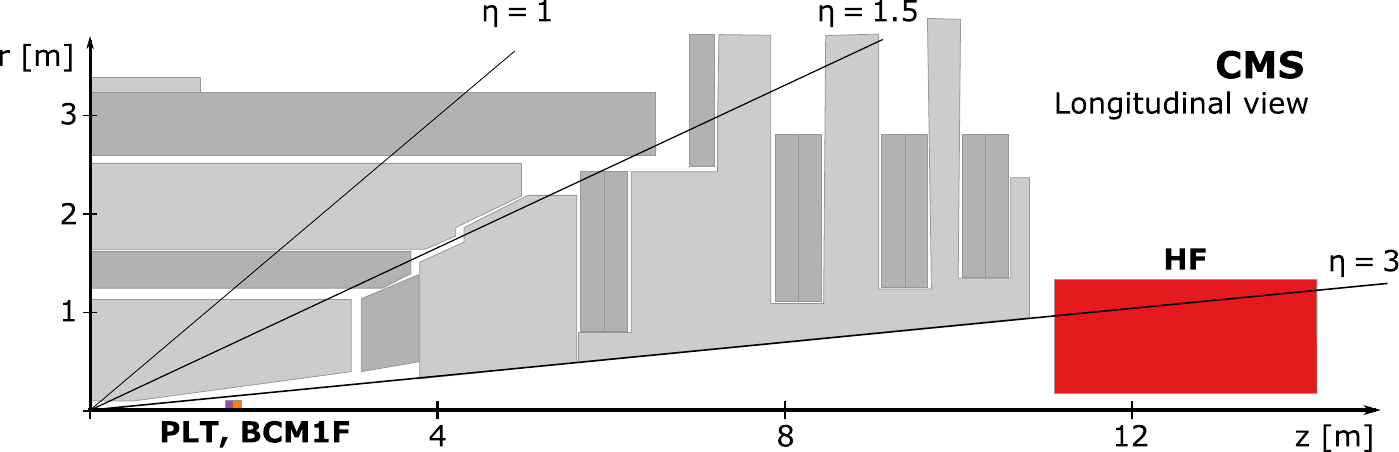}
\caption{%
    Schematic cross section through the CMS detector in the $r$--$z$ plane.
    The luminometers used during \PbPb collisions in 2015 and 2018 are highlighted.
    The approximate locations of PLT, BCM1F, and HF with respect to the IP are indicated with purple, orange, and red rectangles, respectively.
    The center of the detector, corresponding to the approximate position of the \PbPb collision point, is located at the origin.
    The parts of the muon detectors at radii larger than 4\unit{m} are omitted in this schema.
}
\label{fig:luminometers}
\end{figure*}

The PLT, HF, and BCM1F data, as well as beam-related LHC data, are collected and stored in a time-based rather than an event-based manner by using a dedicated DAQ system that is independent of the central CMS DAQ system~\cite{Khachatryan:2016bia, Tapper:1556311}.
Brief descriptions of the three luminometers and the respective algorithms used to estimate the instantaneous luminosity are discussed in the following.
An overview of the position of these luminometers within the CMS detector is shown in Fig.~\ref{fig:luminometers}.

\subsection{Pixel luminosity telescope}

The PLT is a dedicated system for measuring luminosity using silicon pixel sensors, which was installed at the beginning of 2015~\cite{CMSBRIL:2022qin}.
There are a total of 48 sensors arranged into 16 ``telescopes'', eight at either end of CMS outside the pixel endcap, corresponding to $\abs{\eta}\approx4.2$.
Each telescope contains three sensor planes arranged in a triplet that faces the IP.
The sensors measure $8{\times}8\unit{mm}^2$, and are divided into 80 rows and 52 columns of pixels.
The PLT measures the rate of triple coincidences, where a hit is
observed in all three planes, typically corresponding to a track from a particle originating at the IP.

\subsection{Forward hadron calorimeter}

The HF calorimeters use steel as an absorber and quartz fibers as the sensitive material.
The two halves of the HF are located 11.2\unit{m} from the IP, one on each end, and together they provide coverage in the range $3.0<\abs{\eta}<5.2$.
Both HF calorimeters consist of 432 readout towers, containing long and short quartz fibers running parallel to the beam.
The long fibers run the entire depth of the HF calorimeter (165\unit{cm}, or approximately 10 interaction lengths), whereas the short fibers start at a depth of 22\unit{cm} from the front of the detector.

{\tolerance=800
The HF back-end electronics consist of two independent readout systems:\ a primary readout over optical links for physics data, and a secondary readout using Ethernet links, explicitly reserved for luminosity data.
Therefore, the HF luminosity measurement can be performed at the full 40\unit{MHz} bunch crossing rate~\cite{CERN-LHCC-97-031}.
The back-end electronics upgrade during 2013--2015 added new electronics using field-programmable gate array technology such that several features of the readout
were separately programmable for luminosity histogramming, \ie, identifying and counting the readout channels.
Although the whole HF is capable of being read out for luminosity use, only two of the twelve rings
(31 and 32 using the numbering in Ref.~\cite{CMS:PRF-21-001}) in the range $3.15<\abs{\eta}<3.5$ are used to ensure uniform occupancy optimized based on full detector simulation studies.
\par}

For the computation of the HF observable used in the luminosity measurement, the occupancy method (HFOC) is used.
In this method, to limit the contribution of noise, the fraction of channels with signal above an analog-to-digital converter threshold is used for each bunch crossing identification number (BCID) in a configurable time window.
The threshold is set high enough to avoid most noise and as low as possible otherwise.
The number of valid measurements is also stored, so the fraction of events with hits above threshold can be computed.

\subsection{Fast beam conditions monitor}

The BCM1F measures luminosity and beam-induced background~\cite{Bell_2010}.
It consists of a total of 24 sensors mounted on the four carriages that also hold the PLT
planes, providing measurements at both sides of the IP, corresponding to $\abs{\eta}\approx4.1$.
The BCM1F installed in 2014 used single-crystalline diamond sensors.
Radiation damage necessitated their exchange in 2017 with a mixture of single-crystalline diamond, poly-crystalline diamond, and silicon sensors.
The diamond sensors used split-pad metallization, with each sensor having two readout channels to keep the overall occupancy low given the expected conditions in 2015--2018.

The BCM1F features a fast readout with 6.25\unit{ns} time resolution.
The precise time measurement, facilitated by the position of BCM1F at 1.8\unit{m} from the IP, allows hits from outgoing collision products to be separated from incoming beam-induced background hits.
The number of hits in the poly-crystalline diamond sensors is used as the BCM1F observable for the 2018 luminosity measurement.

\subsection{Interaction rate determination}

The interaction rate $R(t)$ in Eq.~\eqref{eq:lumirel} is measured with the PLT, HFOC, and BCM1F luminometers by
applying the ``zero counting'' method~\cite{CMS:2021xjt} to the respective measured observables.
Specifically, we calculate the fraction $p(0)$ of bunch crossings with no detected interaction, \ie, no particle
traversing through the luminometer. This method avoids potential systematic effects from overlapping tracks being counted as a single hit.
The mean rate is calculated by inverting the Poisson distribution: $R=-\ln p(0)$.
This method remains reliable in low detector occupancy conditions, where the probability of observing zero interactions is high.
In general, it offers better linearity than simple rate counting. Background contributions to the rate measurements are discussed in Section~\ref{sec:bkg}.

\section{Luminosity scan method}
\label{sec:vdmScan}

\subsection{Absolute luminosity calibration}
\label{sec:Calibration}

The instantaneous luminosity of a colliding bunch pair separated by $(\Dx,\Dy)$ in the $x$--$y$ plane transverse to the beam directions is given by
\ifthenelse{\boolean{cms@external}}{\begin{multline}
    \Lumi(\Dx,\Dy)\\
    =\None\Ntwo f\iint\!\rho_1(x,y)\rho_2(x+\Dx,y+\Dy)\,\rd x\,\rd y \\
    =\frac{\None\Ntwo f}{\Aeff(\Dx,\Dy)},
\end{multline}}{\begin{equation}
    \Lumi(\Dx,\Dy)
    =\None\Ntwo f\iint\!\rho_1(x,y)\rho_2(x+\Dx,y+\Dy)\,\rd x\,\rd y
    =\frac{\None\Ntwo f}{\Aeff(\Dx,\Dy)},
\end{equation}}
where \None, \Ntwo are the numbers of particles in the two colliding bunches, $f=11\,246\unit{Hz}$ is the LHC orbit frequency~\cite{Bruning:2004ej}, and $\rho_1$, $\rho_2$ are the particle densities in the two colliding bunches.
The effective transverse beam overlap integral \Aeff can be expressed in terms of the convolved beam widths \Sigx and \Sigy as
\begin{equation}\label{eq:CapSigmadef}
    \Aeff=2\pi\Sigx\Sigy.
\end{equation}
For a reference separation $(\Dxz,\Dyz)$, often chosen as ``head-on'' with zero separation $\Dxz=\Dyz=0$, the convolved beam width in one direction can be obtained by varying the separation in the orthogonal direction and integrating over the rate:
\ifthenelse{\boolean{cms@external}}{\begin{equation}\begin{aligned}
    \Sigx&=\frac{1}{\sqrt{2\pi}R(\Dxz,\Dyz)}\int\! R(\Dx,\Dyz)\,\rd\Dx, \\
    \Sigy&=\frac{1}{\sqrt{2\pi}R(\Dxz,\Dyz)}\int\! R(\Dxz,\Dy)\,\rd\Dy.
\end{aligned}\end{equation}}{\begin{equation}
    \Sigx=\frac{1}{\sqrt{2\pi}R(\Dxz,\Dyz)}\int\! R(\Dx,\Dyz)\,\rd\Dx,
    \quad
    \Sigy=\frac{1}{\sqrt{2\pi}R(\Dxz,\Dyz)}\int\! R(\Dxz,\Dy)\,\rd\Dy.
\end{equation}}
Experimentally, the integration over $\rd\Dx$ and $\rd\Dy$ is implemented in the form of a pair of vdM scans~\cite{vanderMeer:296752, CMS:2021xjt}.
Using the linear relation in Eq.~\eqref{eq:lumirel}, the luminosity measured from a vdM scan pair in this way can then be used to measure \sigmaVis for a given luminometer as
\begin{equation}\label{eq:sigVis}
    \sigmaVis=\frac{2\pi\Sigx\Sigy R(\Dxz,\Dyz)}{\None\Ntwo f}.
\end{equation}
All quantities on the right-hand side of Eq.~\eqref{eq:sigVis} are measurable.
In particular, \Sigx and \Sigy are determined from fits to the luminometer rate as a function of the beam separation (``scan distributions'') for vdM scans in $x$ and $y$, respectively, and $R(\Dxz,\Dyz)$ is taken as the mean of the fitted amplitudes from the two scans. More details and an example are given in Section~\ref{sec:ExpSetup} and Fig.~\ref{fig:PLTXYFits_Fill4689}.
We evaluate \sigmaVis separately for each BCID, and separately for vdM scan pairs that are performed sufficiently close in time to minimize the impact of the emittance evolution of the bunches.

The parameters \Sigx and \Sigy represent physical dimensions of the colliding bunches, and measurements with different luminometers are thus expected to yield the same result.
The comparison of the values obtained with different luminometers thus represents a useful cross-check.
In contrast, $R(\Dxz,\Dyz)$ and hence \sigmaVis differ depending on the active area and position of the employed luminometer.

In Eq.~\eqref{eq:CapSigmadef}, the vdM scan method assumes that \Sigx and \Sigy are independent of each other, \ie, that the $\rho_i$ of the two beams factorize in $x$ and $y$:
\begin{equation}\label{eq:factorization}
    \rho_i(x,y)=\rho_{i,x}(x)\rho_{i,y}(y).
\end{equation}
The estimate of the bias introduced by this assumption is given in Section~\ref{sec:xyCor}.

\subsection{Experimental setup}
\label{sec:ExpSetup}

\begin{figure*}[!htp]
\centering
\includegraphics[width=\textwidth]{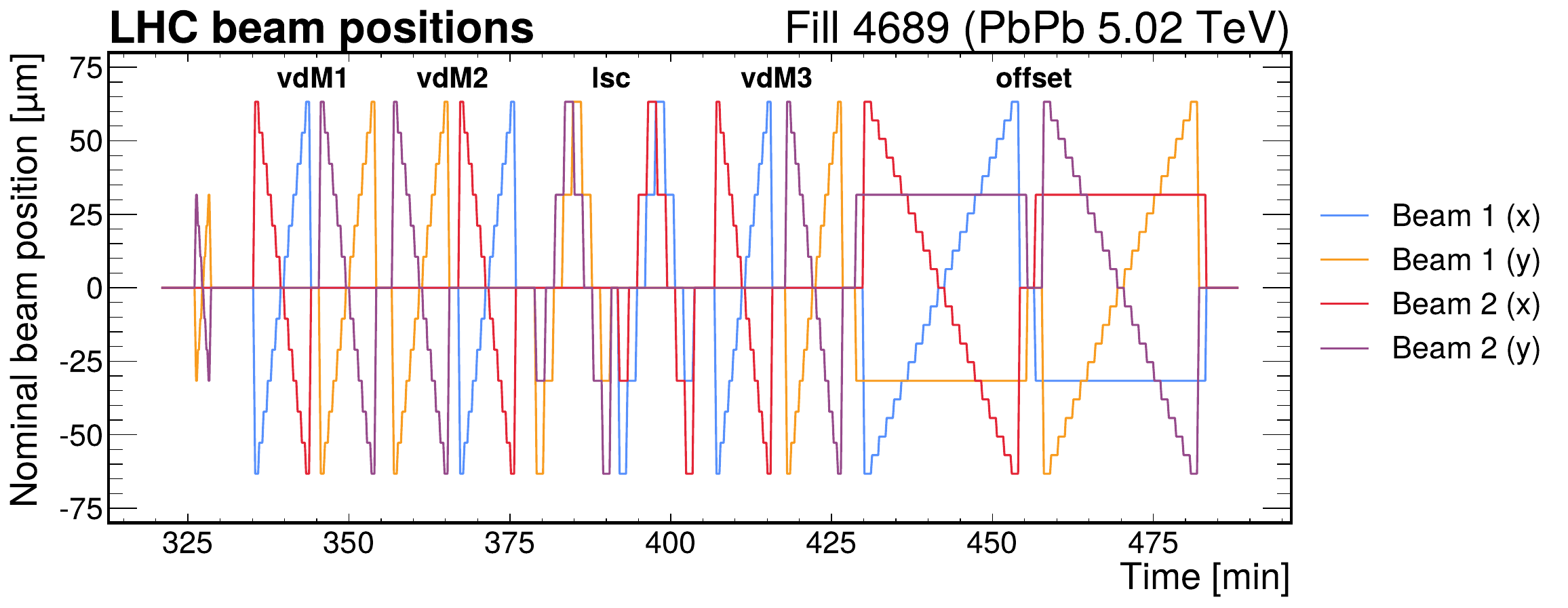} \\[1ex]
\includegraphics[width=\textwidth]{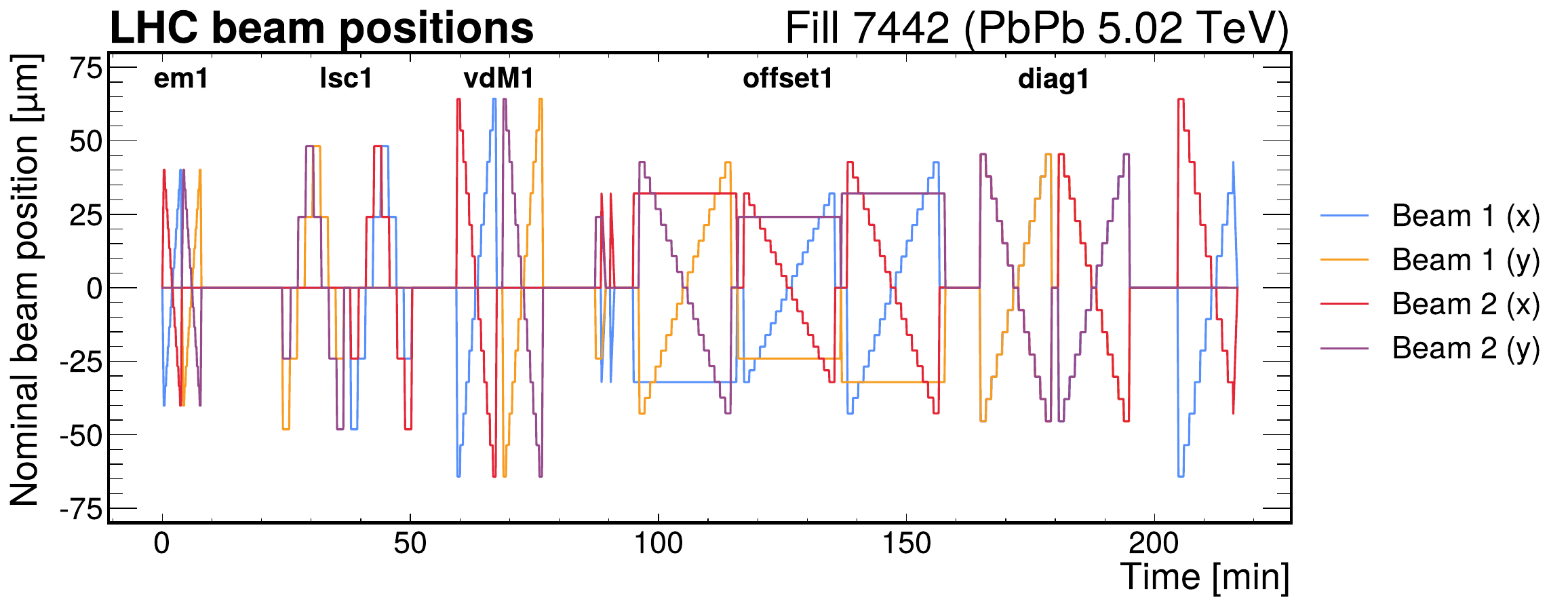} \\[1ex]
\includegraphics[width=\textwidth]{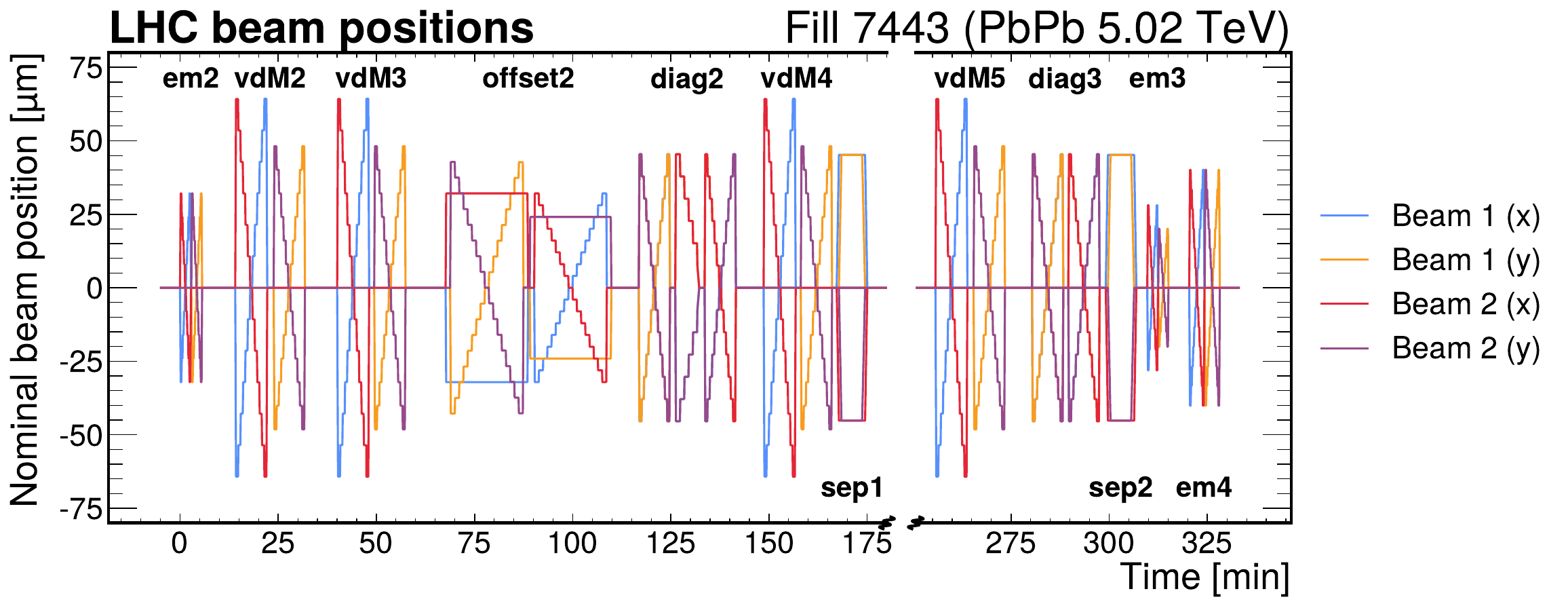}
\caption{%
    Horizontal and vertical beam displacements derived from the LHC corrector magnet currents during the vdM scan program in fill 4689 (upper), fill 7442 (middle), and fill 7443 (lower).
    The first $x$ offset scan in fill 7442 was performed with smaller then planned maximal separation and was retaken.
    The last vdM scan in fill 7442 was interrupted by a premature beam dump and is thus not used in the analysis.
}
\label{fig:beamPositions}
\end{figure*}

{\tolerance=800
The 2015 \PbPb luminosity scan program was performed in the LHC fill 4689 in December 2015, and the 2018 program in fills 7442, 7443, and 7483 in November 2018.
The LHC conditions during these four fills are summarized in Table~\ref{tab:lhcconditions}.
The bunch pairs colliding at IP5 were widely spread over the orbit, with a minimum separation between the bunch crossings of 225 (150)\unit{ns} for 2015 (2018).
The LHC beam optics at IP5 was adjusted to $\betast=0.8$ (0.5)\unit{m} and a half horizontal crossing angle $\phi$ of 0 (160)\unit{$\mu$rad}.
The beams were injected with an expected transverse bunch size at IP5 of $\sigmab=27\mum$ in the $x$ direction and of $\sigmab=20\mum$ in the $y$ direction.
The beam intensities were measured with the DC current transformers (DCCTs)~\cite{Barschel:1425904}, and the individual bunch currents with the fast beam current transformers (FBCTs)~\cite{Belohrad:1267400}.
The longitudinal density monitors (LDMs)~\cite{PhysRevSTAB.15.032803} are used for the estimation of ghost and satellite charges, as described in Section~\ref{sec:spurious}.
The beam orbit during the scan program was monitored using two beam position monitor (BPM) devices:\ the diode orbit and oscillation system (DOROS)~\cite{Gasior:1476070} located near IP5, and the BPMs in the nearby LHC arcs (``arc BPMs'') whose measurements are extrapolated to IP5 using the LHC optics file~\cite{lhc_optics}.
The beam separation scans performed during these four fills are summarized in Table~\ref{tab:scanprogram}, and described in the following.
In Fig.~\ref{fig:beamPositions}, the beam positions as a function of time are shown for fills 4689, 7442, and 7443, illustrating the scan program.
\par}

\begin{table*}[!ht]
\centering
\topcaption{%
    Summary of the LHC conditions at IP5 for the 2015 and 2018 \PbPb luminosity scan programs.
    The table shows the number $n_{\text{b}}$ of colliding bunch pairs, $\phi$, \betast, the targeted proton-equivalent transverse emittance $\varepsilon$ at injection, the beam intensities \None, \Ntwo, and the initial instantaneous luminosity $\Lumi_{\text{init}}$.
}
\label{tab:lhcconditions}
\renewcommand\arraystretch{1.2}
\begin{tabular}{cccccccc}
    \hline
    \multirow{2}{*}{Year} & \multirow{2}{*}{Fill} & \multirow{2}{*}{$n_{\text{b}}$} & $\phi$ & \betast & $\varepsilon$ & {\None} / \Ntwo & $\Lumi_{\text{init}}$ \\[-1pt]
    & & & [$\mu$rad] & [m] & [$\mu$m] & [$10^{11}$ particles] & [$10^{27}\percms$] \\
    \hline
    2015 & 4689 & 400 & 0 & 0.8 & 2.0 & 43.0 / 44.7 & 1.25 \\[\cmsTabSkip]
    2018 & 7442 & 288 & 160 & 0.5 & 3.5 & 97.3 / 98.7 & 2.11 \\
    & 7443 & 288 & 160 & 0.5 & 3.5 & 91.2 / 92.8 & 1.91 \\
    & 7483 & 620 & 160 & 0.5 & 3.5 & 96.6 / 98.5 & 3.98 \\
    \hline
\end{tabular}
\end{table*}

\begin{table}[!ht]
\centering
\topcaption{%
    Summary of the beam separation scans performed during the 2015 and 2018 \PbPb luminosity scan programs.
    The table shows the total number of vdM, length scale calibration (``lsc''), off-axis (offset or diagonal), and emittance (``em'') scan pairs that were performed, as well as the number of super-separation periods (``sep'').
    Only the scans used for the measurements reported here are included.
}
\label{tab:scanprogram}
\renewcommand\arraystretch{1.1}
\begin{tabular}{ccccccc}
    \hline
    Year & Fill & vdM & lsc & off-axis & em & sep \\
    \hline
    2015 & 4689 & 3 & 1 & 1 & 0 & 0 \\[\cmsTabSkip]
    2018 & 7442 & 1 & 1 & 2 & 1 & 0 \\
    & 7443 & 4 & 0 & 3 & 3 & 2 \\
    & 7483 & 0 & 1 & 0 & 1 & 0 \\
    \hline
\end{tabular}
\end{table}

In fill 4689, 10 scans were recorded:\
one \XY vdM scan pair (``vdM1''),
one \YX vdM scan pair (``vdM2''),
one \YX length scale calibration scan pair (``lsc''),
one more \XY vdM scan pair (``vdM3''),
and one \XY scan pair with constant beam separation in the nonscanning direction (``offset'').
For the vdM scans, the maximum transverse separation between the two beams was about $\pm5$ (6)\usigmab in $x$ ($y$), and the scans were performed in a sequence of 13 steps with a step size of 0.5\usigmab.
During the length scale calibration scans, the beams were separated by a fixed amount of about 32\mum in one transverse direction, corresponding to 1.2 (1.6)\usigmab in the $x$ ($y$) direction, and then moved coherently forward and backward in four steps in the same transverse direction.
At each scan point, the rates are integrated for 30\unit{s}.
In the offset scans, beams were kept at the fixed separation of about 32\mum in one coordinate while the other coordinate was separated and scanned in 22 steps from $-63$ to $+63\mum$, corresponding to about 6\usigmab total distance, with an integration time of 61\unit{s} per scan point.

{\tolerance=800
In fill 7442, 11 scans were recorded:\
one \XY emittance scan pair (``em1''),
one \YX length scale calibration scan pair (``lsc1''),
one \XY vdM scan pair (``vdM1''),
a set of one $y$ and two $x$ scans with a constant separation in the nonscanning direction (``offset1''),
and one diagonal scan pair with first opposite-sign and then equal-sign separation in the $x$ and $y$ directions (``diag1'').
Due to a premature beam extraction (``dump'') out of the accelerator ring, a following vdM scan pair was interrupted during the $x$ scan, and the scan program was continued in the consecutive fill.
In fill 7443, 13 scans were recorded in the first part of the program:\
one \XY emittance scan pair (``em2''),
two \XY vdM scan pairs (``vdM2'' and ``vdM3''),
one \YX pair with a constant separation in the nonscanning direction (``offset2''),
one diagonal scan pair (``diag2''),
one more \XY vdM scan pair (``vdM4''),
and a period with a large beam separation (``super separation'', ``sep1'').
To probe reproducibility of the \sigmaVis calibration, a second part with 5 different types of scans was recorded later during the same fill:\
one \XY vdM scan pair (``vdM5''),
one diagonal scan pair (``diag3''),
a super-separation scan (``sep2''),
and two \XY emittance scan pairs (``em3'' and ``em4'').
In fill 7483, the reproducibility of the length scale calibration was tested with a similar scan pattern as shown for fill 7442, and is discussed in Section~\ref{sec:lsc}.
\par}

For the 2018 vdM scans, the two beam orbits were separated up to about $\pm$5 (6)\usigmab in $x$ ($y$),
and scanned in 13 steps across one another with an integration time of 30\unit{s} per scan point.
Emittance scans covered a smaller orbit movement of about 2--3\usigmab and had only 10\unit{s} per scan point.
In the offset scans, 17 steps up to an orbit movement of $\pm$3\usigmab were performed while separating the two beams by about 2 or 2.5\usigmab in the nonscanning direction.
In the diagonal scans, beam orbits were moved by up to $\pm$3\usigmab and simultaneously scanned in 13 steps across one another.
The integration time at each scan point was 60\unit{s} for both offset scans and the diagonal scans in fill 7442, and 30\unit{s} for the diagonal scans in fill 7443 to probe the impact of orbit drifts.
For the length scale calibration scans, the two beams are kept at a fixed separation of about 1\usigmab and moved coherently forward and backward in four steps, with about 73\unit{s} integration time per scan point.
The super-separation periods of 5\unit{min} each have the beams separated by 3.3 (4.5)\usigmab in $x$ ($y$).

\begin{figure}[!ht]
\centering
\includegraphics[width=0.47\textwidth]{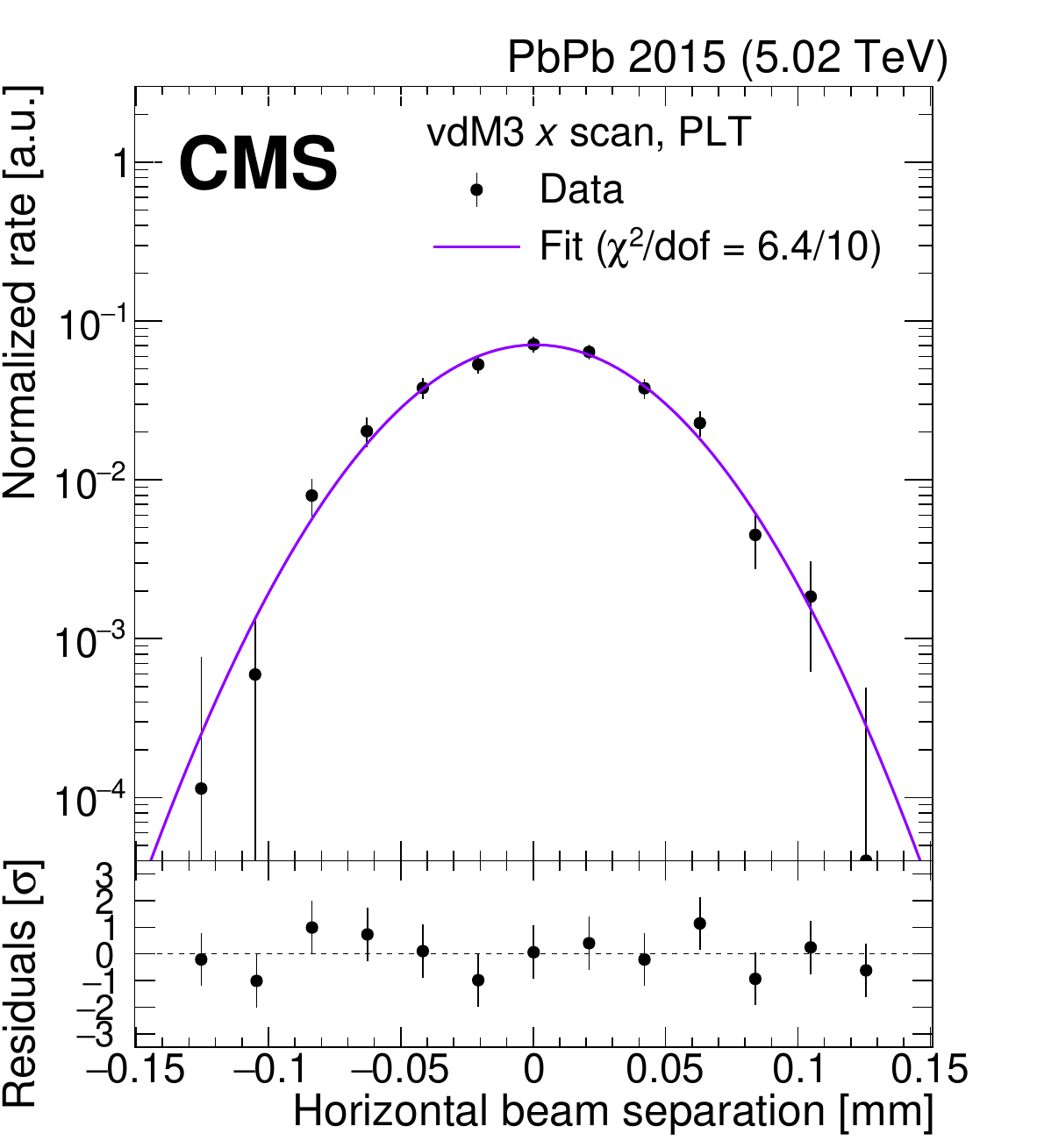}%
\hfill%
\includegraphics[width=0.47\textwidth]{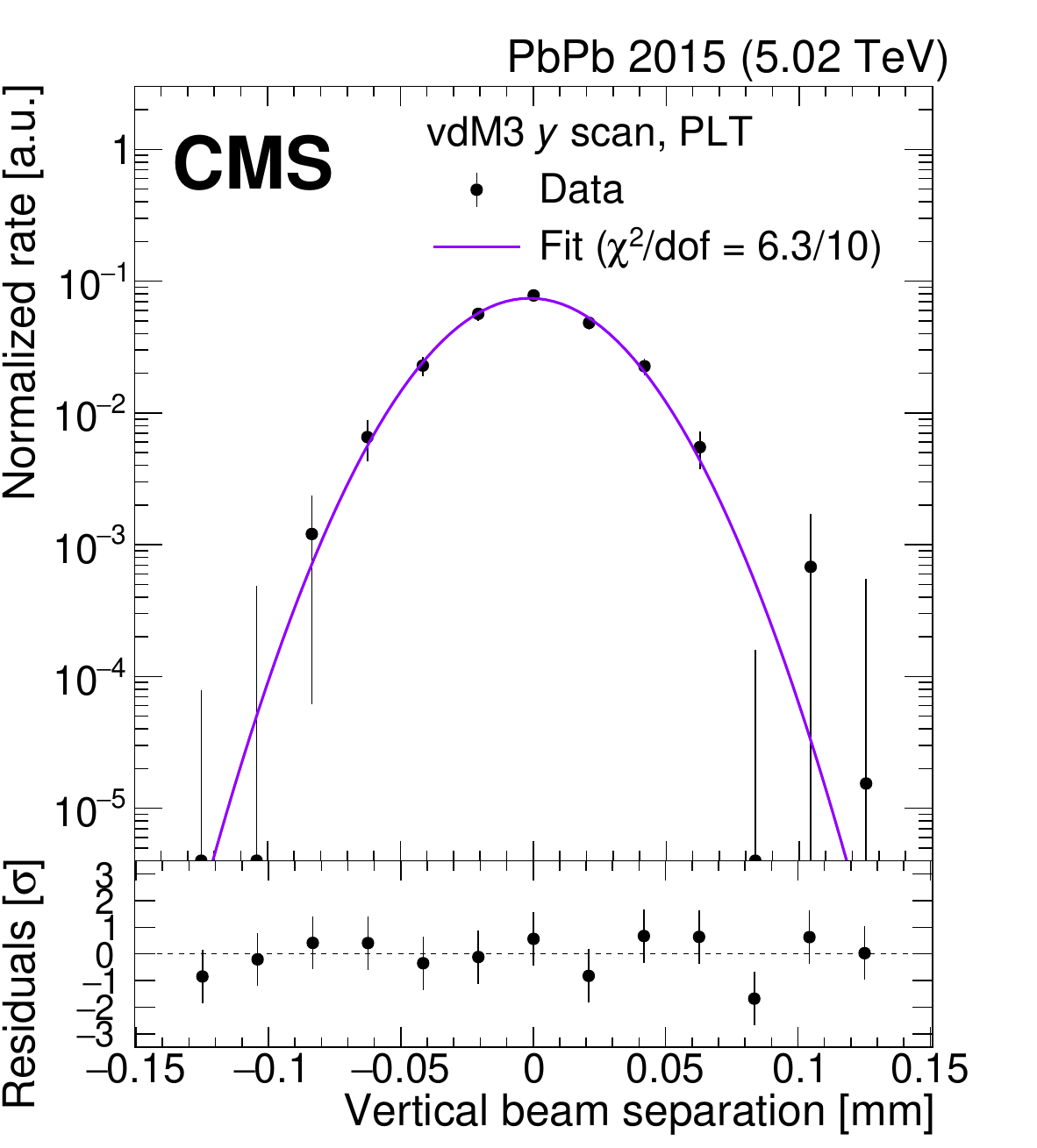}
\caption{%
    Examples of 2015 scan distributions, \ie, the normalized rate recorded by PLT in BCID 955 as a function of the beam separation in $x$ (\cmsLeft) and $y$ (\cmsRight), respectively.
    The error bars represent the statistical uncertainty in the measured rate.
    The purple curve corresponds to the single-Gaussian fit, and the obtained \chisqred is given on the plots.
    The lower panels show the residuals, defined as the difference between the measured and fitted values divided by the statistical uncertainty in the rate.
}
\label{fig:PLTXYFits_Fill4689}
\end{figure}

{\tolerance=1600
The vdM scan distributions are fitted with a single-Gaussian model that is the best performing parametrization based on the goodness-of-fit metric \chisq divided by the number of degrees of freedom (\dof).
Example fits using PLT data for the 2015 vdM3 scan are shown in Fig.~\ref{fig:PLTXYFits_Fill4689}.
\par}

\section{Absolute luminosity calibration corrections}
\label{sec:calibrations}

A precise measurement of the luminosity requires corrections and the evaluation of systematic uncertainties for all ingredients in Eq.~\eqref{eq:sigVis}.
Corrections to \None, \Ntwo are discussed in Section~\ref{sec:spurious}.
In Section~\ref{sec:bkg}, corrections for parasitic contributions to the measured luminometer rates are presented.
Beam separation corrections, relevant for the \Sigx, \Sigy fits, are introduced in Sections~\ref{sec:OD}--\ref{sec:beambeam}.
Bunch shape corrections for the fits are discussed in Section~\ref{sec:beambeam}, and corrections for the factorizability assumption of Eq.~\eqref{eq:factorization} in Section~\ref{sec:xyCor}.

\subsection{Bunch current normalization}
\label{sec:spurious}

{\tolerance=1600
The overall beam current is measured with a high accuracy of 0.2\% by the DCCT~\cite{Barschel:1425904}, whereas individual bunch currents are measured by the FBCT~\cite{Belohrad:1267400} with a lower absolute precision.
To obtain the most precise measurement of the individual bunch currents, the FBCT measurement is normalized to match the DCCT measurement.
Additionally, the FBCT and DCCT measurements need to be corrected for spurious charges, where we consider two potential sources:\ charges belonging to nominally empty (unfilled) BCIDs (``ghosts''); and those within a filled BCID but outside of the nominal time window for collisions (``satellites'').
Corrections for ghost and satellite charges are derived from the LDM measurement, and applied following the methodology described in Ref.~\cite{CMS:2021xjt}.
An independent measurement of the ghost charge provided by the LHCb Collaboration with the beam-gas imaging method~\cite{Barschel:1693671} confirms the LDM results.
The estimated ghost and satellite fractions are listed in Table~\ref{tab:GhostSatCorrections}.
Ghost charges were nonnegligible, especially in 2015, and show a time dependence for both beams that is taken into account.
\par}

\begin{table}[!ht]
\centering
\topcaption{%
    Ghost fractions ($f_{\text{g}}$) and satellite fractions ($f_{\text{s}}$) per beam, and total impact on \sigmaVis ($\Delta\sigmaVis$).
}
\label{tab:GhostSatCorrections}
\renewcommand\arraystretch{1.2}
\cmsTable{\begin{tabular}{ccccccc}
    \hline
    \multirow{2}{*}{Year} & \multirow{2}{*}{Scan} & \multicolumn{2}{c}{Beam 1} & \multicolumn{2}{c}{Beam 2} & $\Delta\sigmaVis$ \\[-2pt]
    & & $f_{\text{g}}$ [\%] & $f_{\text{s}}$ [\%] & $f_{\text{g}}$ [\%] & $f_{\text{s}}$ [\%] & [\%] \\[2pt]
    \hline
    2015 & vdM1 & 4.0 & 0.8 & 2.3 & 0.6 & $+8.2$ \\
    & vdM2 & 4.1 & 0.8 & 2.4 & 0.6 & $+8.3$ \\
    & vdM3 & 4.2 & 0.9 & 2.4 & 0.6 & $+8.6$ \\[\cmsTabSkip]
    2018 & vdM1 & 0.8 & 0.4 & 0.9 & 0.4 & $+2.5$ \\
    & vdM2 & 1.2 & 0.5 & 1.5 & 0.6 & $+3.9$ \\
    & vdM3 & 1.3 & 0.5 & 1.6 & 0.6 & $+4.0$ \\
    & vdM4 & 1.4 & 0.6 & 1.7 & 0.6 & $+4.4$ \\
    & vdM5 & 1.4 & 0.6 & 1.8 & 0.7 & $+4.7$ \\
    \hline
\end{tabular}}
\end{table}

The resulting overall spurious charge correction for fill 4689 is 8.4\%, with a systematic uncertainty of 0.3\% based on the difference between bunched and total (\ie, including charges not captured by the RF buckets and therefore drifting around the LHC ring) spurious charge fractions as measured from LDM.
Similarly, corrections of 2.5 (3.9--4.6)\% are found for fill 7442 (7443) with an uncertainty of 0.5\%.
The DCCT accuracy is included as an additional systematic uncertainty.

\subsection{Noncollision rate}
\label{sec:bkg}

We identify and account for three sources of noncollision background, \ie, rate contributions that are not the direct result of the collision in the given 25\unit{ns} time window:\ (i) beam-induced background (BIB); (ii) intrinsic detector noise; and (iii) out-of-time background due to various effects including electronics time walk and material activation background (``afterglow'').
The noise contribution can be estimated from noncolliding BCIDs where no bunch is present for either beam (``empty'' bunches), whereas the BIB contribution can be estimated when a bunch is present for only one beam (``unpaired'' bunches).
The filling scheme is characterized by three or more empty bunch slots between two filled bunch slots, and the filled bunch slots are grouped in so-called ``trains''.
The out-of-time contribution can thus be estimated from the rate in the empty BCID just before the filled one.
The spurious charge effect impacts all interaction points uniformly and is inherently tied to beam current measurements, whereas noncollision rate arises from sources specific to each detector. Given the large difference in magnitude between these effects, they are treated as uncorrelated
In 2018, the super-separation scans are also used to probe the noncollision rate.
Since the collision rate is negligibly low during super-separation scans, the measured luminometer rate provides an estimate of the noncollision rate.
We study all three luminometers separately, and find the noncollision rate to be detector dependent, impacting $R(\Dxz,\Dyz)$, \Sigx, and \Sigy with a nonnegligible effect in all cases.
The results are described for each luminometer separately below.

\subsubsection{PLT}

The PLT triple coincidence rate in empty BCIDs is nonzero only in BCIDs immediately after collisions, and we thus conclude that the noise and long time constant afterglow contributions are negligible. Thus, the background for PLT primarily originates from out-of-time effects with short ($<$25\unit{ns}) time constant, such as electronic time walk.

For 2015, the rate of unpaired bunches and empty bunches of the abort gap during all vdM scans are used to measure the noncollision rate.
The corresponding systematic uncertainty in \sigmaVis, typically 0.5\%, is evaluated from the difference with subtracting the noncollision rate estimated from the rate of colliding bunches at maximum separation (where the beams are separated by more than 6\usigmab) during the vdM scans.

For 2018, the super-separation scans are used instead.
In the sep1--2 scans, the average rate with the most probable value (``mode'') of 5.3 (2.7)${}\times10^{-7}$ for colliding (unpaired) BCIDs is compatible between the two super-separation periods.
The mode is also considered as the statistical uncertainty in the noncollision rate.
The systematic uncertainty is estimated as the difference between the noncollision rate from unpaired and empty BCIDs in the abort gap during the vdM1--5 scans, impacting \sigmaVis by up to 0.2\%.

The effect of the subtraction of the noncollision rate on \sigmaVis is $-2.5$ ($-0.6$)\% in 2015 (2018).
As the noncollision rate contributes less than 0.1\% of the peak value for PLT, this value is used as the systematic uncertainty due to noncollision rate during routine \PbPb data-taking conditions.

\subsubsection{HFOC}

The head-on rate from empty BCIDs in and out of trains during vdM scans revealed out-of-time contribution features.
The rate immediately after collisions is higher than the average baseline in between two colliding BCIDs, and it slowly decreases in the other empty BCIDs.
Therefore, the rate from the empty BCID before collisions corresponds to out-of-time contribution and, to a lesser extent, noise and BIB.
The rate from unpaired BCIDs is not significantly higher than that from empty BCIDs.
We thus consider the rate from the empty BCID before the colliding one as the total background.

For the measured HFOC rates, the background is relatively large with values of up to 4 (10)\% of the peak rate in 2015 (2018).
The impact of this correction on \sigmaVis is found to be $-8\%$ in 2015, whereas the \chisq minimization using the single-Gaussian fit as described in Section~\ref{sec:ExpSetup} for 2018 does not converge without this background subtraction.
We apply a cross-check method in 2018 as described in Section~\ref{sec:stability}, in which we iteratively parametrize the rate, and find agreement within 0.5\%,
which is assigned as the systematic uncertainty due to the noncollision rate.

\subsubsection{BCM1F}

Corrections accounting for the out-of-time contribution are applied online to BCM1F data in 2018, and thus the head-on rate from empty BCIDs in and out of trains during vdM scans does not display significant out-of-time contribution features in the offline data.
The noncollision rate is estimated as the average rate from colliding BCIDs in the sep1--2 scans, and amounts to 0.2\% of the head-on rate.

The subtraction of the noncollision rate changes \sigmaVis by $-3.7\%$.
The difference between the noncollision rate from unpaired and
empty BCIDs in the abort gap during the vdM1--5 scans is assigned as systematic uncertainty, impacting \sigmaVis by up to 1.0\%.

\subsection{Orbit drift}
\label{sec:OD}

Orbit drift at the IP during a scan session can impact the beam separations used in the scan distributions.
Two corrections are applied to account for a possible bias.
First, a linear orbit drift correction is applied that corrects for slow orbit drifts and is evaluated as a linear function of time between head-on periods before, during, and after scans.
Second, residual orbit drifts are corrected separately at each scan step, by evaluating the difference between the measured beam positions and those evaluated from the LHC orbit corrector magnet currents with all corrections for known effects applied.
Possible sources for residual orbit drift are random orbit drifts beyond the linear approximation (\eg, a sudden movement during a single step), as well as systematic deviations caused, \eg, by hysteresis of the LHC orbit corrector magnets.
The latter would affect most scan pairs in a similar way, and the corresponding bias would thus not appear as a deviation in the scan-to-scan agreement.
Both corrections are derived from beam position measurements provided by the DOROS and arc BPMs.

Linear orbit drift in the scanning direction is evaluated with linear interpolations between a point before a scan to the head-on point in the middle of the scan, and similarly between the head-on point to a point after the scan.
These interpolations are applied as corrections to the beam separations in the scan distributions.
Additionally, linear orbit drift in the nonscanning direction is applied as a rate correction, using the center values and \Sigx, \Sigy values from the fit with the linear orbit drift in the scanning direction already applied.

In fill 4689, the DOROS data show a large drift in the horizontal direction of beam 2, in contradiction with the measurements of the arc BPMs.
Considering that the DOROS system was still in a commissioning phase in 2015, we conclude that the reliability of the DOROS data in this period was compromised.
Thus, we only use the arc BPM data to derive the linear orbit drift correction.
The impact of the linear orbit drift correction in the scanning direction on \sigmaVis is 0.5\%, and the full size of the correction is assigned as systematic uncertainty.
The impact of the linear orbit drift correction in the nonscanning direction is on average 0.1\%.

For the November 2018 scans, the orbit drift is generally less than 5 and 2\mum in the $x$ and $y$ directions, respectively, as measured by the arc BPMs and consistent with the DOROS measurements.
The corresponding correction in \sigmaVis is up to $-0.1\%$ for the scanning direction, and a systematic uncertainty of 0.1\% is assumed.
In the nonscanning direction, the impact is on average 0.1\%.

\begin{figure}[!h]
\centering
\includegraphics[width=0.47\textwidth]{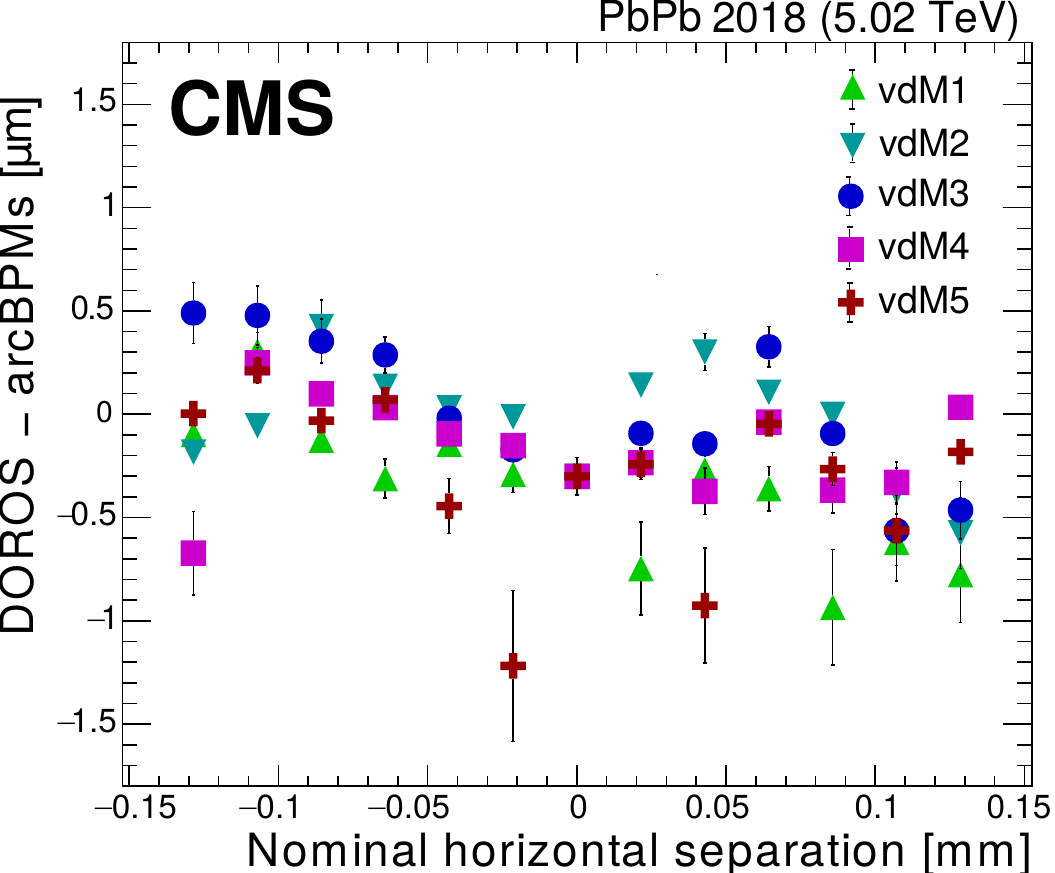}%
\hfill%
\includegraphics[width=0.47\textwidth]{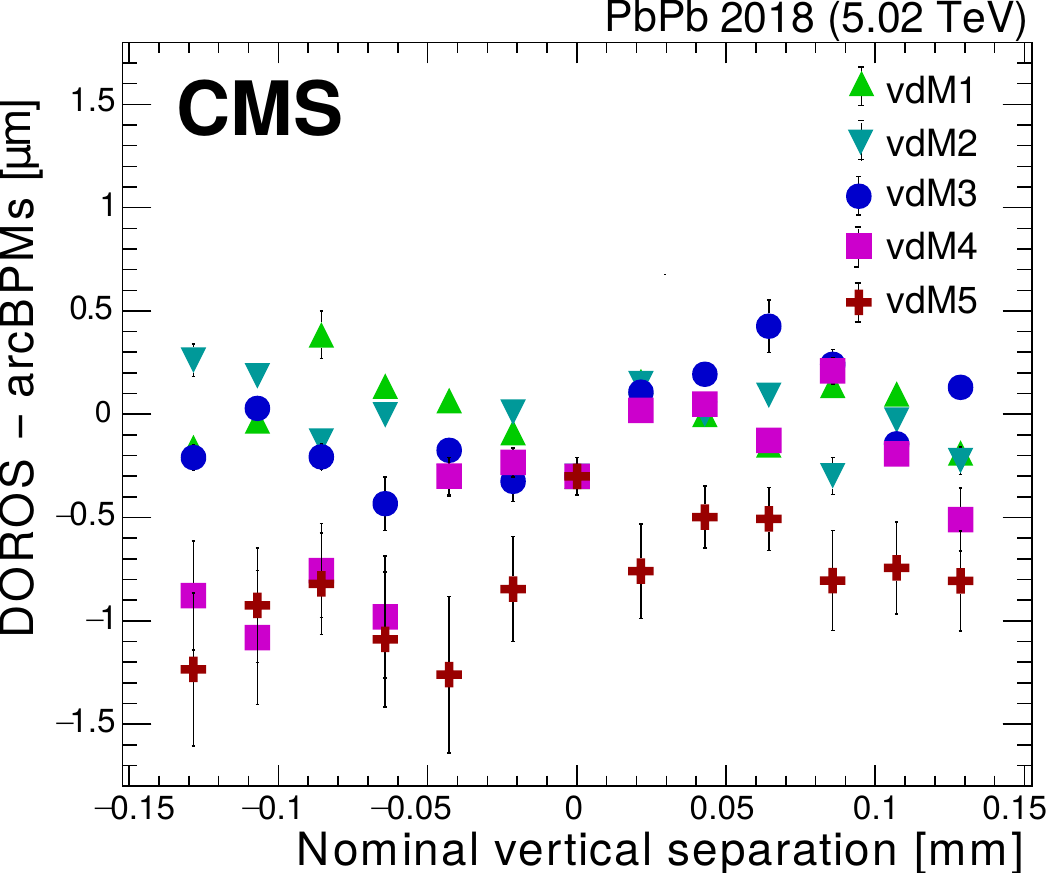}
\caption{%
    The absolute difference in the displacement of the two beams at IP as measured by the DOROS and arc BPMs during the 2018 vdM1--5 scans as a function of the nominal beam separation in $x$ (\cmsLeft) and $y$ (\cmsRight).
    The error bars indicate the statistical uncertainty in the measured beam positions.
 \label{fig:hysteresis}}
\end{figure}

The residual orbit drift correction is evaluated similarly to the procedure described in Ref.~\cite{CMS:2021xjt}.
The beam position at each scan step is evaluated from the DOROS and arc BPM measurements, and corrected for the nominal displacement, beam-beam effects (as described in Section~\ref{sec:beambeam}), length scale per beam and direction, and linear orbit drift.
Owing to $\betast=0.5\unit{m}$, the impact from beam-beam deflection at the location of the DOROS BPMs (21.5\unit{m} away from IP5) is magnified by factors of about 56 and 60 in $x$ and $y$, respectively.
Additionally, we multiply the beam-beam deflection with a factor of 0.46, corresponding to the fraction of colliding bunches at IP5 with respect to all filled bunch slots.
The effect of residual orbit drift on the calibration is then evaluated by considering the difference between the measured and nominal beam separations, resulting in a correction at each scan step.
In 2015, the corrections obtained with the arc BPM measurement results in a change of \sigmaVis of the order of 0.2\%.
Since the DOROS measurement cannot be used for comparison, no correction is applied for the nominal result and the full size of the effect is assigned as a systematic uncertainty.
In 2018, the resulting impact on \sigmaVis is less than 0.1\%, and hence is neglected.
The difference between the results obtained with the DOROS and arc BPMs is shown in Fig.~\ref{fig:hysteresis} for the 2018 vdM scans.
In most cases, the two results differ by less than 1\mum, corresponding to an uncertainty of 0.2\% in \sigmaVis.

\subsection{Length scale calibration}
\label{sec:lsc}

To study the accuracy of the beam separation derived from the LHC corrector magnet currents, we compute a correction that is determined by measuring the luminous region (``beamspot'') movements using the CMS tracking system at the level of micrometer accuracy.
The data recorded during vertical and horizontal length scale calibration scans, described in Section~\ref{sec:ExpSetup} and split into ``forward'' and ``backward'' parts, are used.
In addition, data from the reproducibility fill 7483 are included for 2018, for which the LHC optics setup is identical to the beam configuration in fills 7442--7443 (Table~\ref{tab:lhcconditions}), to constrain a possible fill-dependent bias and study the stability of the derived length scale values.

\begin{figure}[!ht]
\centering
\includegraphics[width=0.47\textwidth]{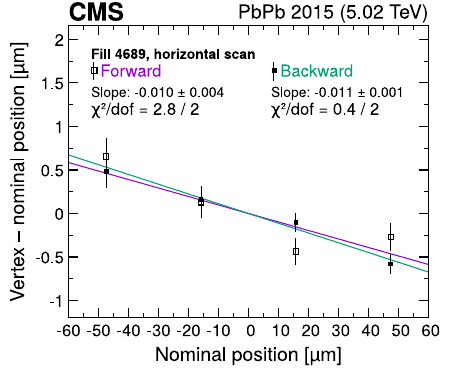}%
\hfill%
\includegraphics[width=0.47\textwidth]{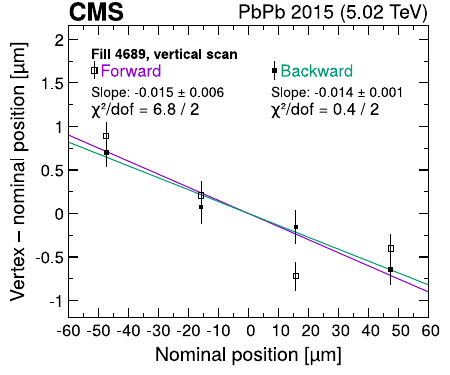}
\caption{%
    Difference between the reconstructed beamspot and the nominal positions as a function of the latter during the 2015 length scale calibration scans.
    The results are shown for the $x$ (\cmsLeft) and $y$ (\cmsRight) scans, separately for the forward (purple) and backward (green) directions.
    The error bars represent the statistical uncertainty in the reconstructed beamspot position.
    The lines denote linear fits with slope and \chisqred values shown in the legend.
}
\label{fig:LSC2015}
\end{figure}

The procedure to derive the length scale constants applies the primary-vertex selection and method described in Ref.~\cite{CMS:2021xjt}.
The beamspot movement as a function of the nominal offset of the beam centroid is fitted with a first-order polynomial, and the calibration constants are extracted in $x$ and $y$ as the magnitude of the fitted slopes.
The measurement is illustrated in Fig.~\ref{fig:LSC2015} for 2015 data, showing the difference between the reconstructed beamspot and the nominal beam positions as a function of the latter such that the fitted slopes represent the length scale correction factors.
It demonstrates an excellent agreement between the results for forward and backward moving scans but large statistical uncertainties for the forward direction on the fitted slope, which is an observed feature of the data.

{\tolerance=800
In nucleus-nucleus collisions, the number of produced particles depends strongly on the geometrical overlap
of the ions at the time of the collision, called ``centrality''~\cite{CMS:2024krd}, which is zero in head-on collisions,
whereas higher values are obtained in the case of peripheral collisions.
The number of charged particles produced per unit of $\eta$ varies from event to event (from few hundred for peripheral collisions to few thousand for the most central collisions) and
the large particle multiplicity of head-on collisions leads to high tracker occupancy.
No effect on the calibration constants is found after repeating the measurement in different bins of centrality.
\par}

The impact on \sigmaVis due to the length scale correction is $-2.4$ ($-1.5$)\% in 2015 (2018).
The uncertainty in \sigmaVis due to the length scale correction is 0.5\% in both years.
In 2015, the uncertainty is dominated by the statistical uncertainties of the fits.
In 2018, the uncertainty contributions are statistical (0.4\%), the difference between the averaged values in fills 7442 and 7483 (0.2\%), and the average impact from residual orbit drifts as estimated using arc BPMs (0.2\%).
In 2015, the higher statistical uncertainty effectively accounted for potential systematic contributions, which were later disentangled in 2018 with a larger sample size.

\subsection{Beam-beam effects}
\label{sec:beambeam}

{\tolerance=800
The strength of the mutual electromagnetic forces of the two colliding positively charged bunches changes as a function of beam separation and, if uncorrected, can bias the results obtained from the vdM scans.
The two beams repel each other (``beam-beam deflection'') changing the actual beam separation during the scans.
A correction to the beam separation in the scan distributions is derived following the procedure described in Refs.~\cite{Kozanecki:1581723, Babaev_2018}, separately per scan and BCID.
In 2015 (2018), an overall impact of less than $+0.1\%$ (between $+0.1$ and $+0.2\%$) was estimated.
In addition to the beam-beam deflection, the electromagnetic forces distort the bunch particle density distributions (often called optical or ``dynamic-$\beta$'' effect), depending on the beam separation.
This effect is anticorrelated to the repulsion and expected to be equally small in \PbPb periods, owing to the low bunch intensities.
No correction is applied, therefore.
A conservative uncertainty due to beam-beam effects is assigned that equals twice the average impact of the beam-beam
deflection, \ie, 0.2 (0.3)\% in 2015 (2018).
\par}

\subsection{Transverse factorizability of the bunch particle density}
\label{sec:xyCor}

As described in Eq.~\eqref{eq:factorization}, the vdM scan method assumes that the bunch particle densities of the two beams are factorizable into $x$- and $y$-dependent components.
This assumption may not hold at the required precision, and result in a biased estimate of \Aeff when using Eq.~\eqref{eq:CapSigmadef}.
This effect can be probed by constructing two-dimen\-sional (2D) luminosity distributions from a combination of vdM (``on-axis'') and offset or diagonal (``off-axis'') scans.
By combining scans that take place close in time, the impacts of both emittance evolution and orbit drift are minimized.
Models of the luminosity distribution that incorporate more complex features, \eg, the presence of a linear transverse correlation or a non-Gaussian distribution, can then be probed with fits of appropriately chosen 2D analytic functions.

The measurement thus makes use of sampling the luminometer rates in the 2D space of the beam separations in $x$ and $y$.
In 2015, this rate distribution is acquired by the combination of the vdM3 and offset scans, whereas in 2018 several combinations of on- and off-axis scans are used as input.
All considered combinations are listed in Table~\ref{tab:xycorr:combinations}.

\begin{table}[!ht]
\centering
\topcaption{%
    Scan combinations used in the transverse factorizability evaluation.
    The components of the scan combinations are listed in time order.
}
\label{tab:xycorr:combinations}
\renewcommand\arraystretch{1.1}
\begin{tabular}{ccr@{~+~}l}
    \hline
    Year & Fill & \multicolumn{2}{c}{Combination} \\
    \hline
    2015 & 4689 & vdM3 & offset \\[\cmsTabSkip]
    2018 & 7442 & vdM1 & offset1 \\
    & & vdM1 & diag1 \\[\cmsTabSkip]
    & 7443 & vdM3 & offset2 \\
    & & vdM3 & diag2 \\
    & & offset2 & vdM4 \\
    & & diag2 & vdM4 \\
    & & vdM5 & diag3 \\
    \hline
\end{tabular}
\end{table}

{\tolerance=1600
The luminometer rates are corrected for background, as well as ghost and satellite contributions.
In 2017, the beam separations are length-scale-calibrated (Section~\ref{sec:lsc}) and corrected for orbit drift per scan point independently in $x$ and $y$ using the arc BPM measurements (Section~\ref{sec:OD}).
No such corrections are applied in 2015, and it was validated that this does not significantly affect the factorizability results.
The length scale cancels out in this correction, and the size of a potential orbit drift is estimated from the rate data and its effect is included as an uncertainty as described below.
\par}

To estimate the size of a residual orbit drift between the constituent on- and off-axis scans, the normalized rates at the intersection point of the vdM and offset scans are used.
The best offset value is evaluated by minimizing the rate difference using a \chisq metric over all the bunch crossings.
The extra offset is estimated to be less than 2\mum for all scan combinations in both years.
The values of the offset agree well when derived using the different luminometer rates, and are used to define the uncertainty in the knowledge of the off-axis scan orbit positions with respect to the vdM scan (``residual misalignment'').

The rate data of the vdM+offset and vdM+diagonal pairs are fitted independently for all calibrated subdetectors and separately for each BCID with a number of 2D analytic models.
The considered models are a single-Gaussian function including linear correlations (``SG''), a sum of two such single Gaussians called a double Gaussian (``DG''), a double Gaussian with common means (``DG2''), and a super-Gaussian (``supG'') model.
The latter is equipped with an additional parameter $p$ with respect to a single-Gaussian function that governs the sharpness of the peak:
\ifthenelse{\boolean{cms@external}}{\begin{multline}
    \text{supG}(x,y)=\frac{A}{\pi\sigma_x\sigma_y\sqrt{1-\varrho^2}}\frac{p}{2^{1/p}\Gamma(p^{-1})} \\
    \times\exp\bigg\{-\frac{1}{2(1-\varrho^2)^p} \bigg[\bigg(\frac{x-x_0}{\sigma_x}\bigg)^2+\bigg(\frac{y-y_0}{\sigma_y}\bigg)^2 \\
    -\frac{2\varrho (x-x_0)(y-y_0)}{\sigma_x\sigma_y}\bigg]^p\bigg\}.
\end{multline}}{\begin{multline}
    \text{supG}(x,y)=\frac{A}{\pi\sigma_x\sigma_y\sqrt{1-\varrho^2}}\frac{p}{2^{1/p}\Gamma(p^{-1})} \\
    \times\exp\bigg\{-\frac{1}{2(1-\varrho^2)^p} \bigg[\bigg(\frac{x-x_0}{\sigma_x}\bigg)^2+\bigg(\frac{y-y_0}{\sigma_y}\bigg)^2
    -\frac{2\varrho (x-x_0)(y-y_0)}{\sigma_x\sigma_y}\bigg]^p\bigg\}.
\end{multline}}
Here, $A$ is the normalization parameter corresponding to the volume under the luminosity distribution, $x_0$ and $y_0$ are the coordinates of the center of the distribution, $\sigma_x$ and $\sigma_y$ are the width parameters, $\varrho$ describes the linear correlation, and $\Gamma$ is the gamma function.
In studying the 2015 data, the same fit models are used with or without an extra constant added, as the presence of a free constant increases the consistency of the fit parameters between PLT and HFOC.
Considering the fitted shapes as the true bunch overlap distributions, simulations of on-axis scans are performed to imitate the vdM measurement.
A correction is derived from the observed biases averaging several pseudo-experiments.
The bias is $-0.4\%$ in 2015 and within 0.9--1.3\% depending on the scan combination in 2018.
These values are averaged over the studied BCIDs with the uncertainty in the mean below 0.1\%.

Various sources of systematic uncertainties are considered.
The residual misalignment is included by shifting the offset scans either orthogonal or along the scan direction by $\pm$2\mum.
Repeating the fit and the pseudo-experiments yields an effect of 0.1\% in 2015 and up to 0.3\% in 2018.

The factorizability correction is evaluated as the average of the obtained values for all calibrated luminometers and all scan combinations.
In 2015, a single vdM and offset scan combination is available, and the difference between the results obtained by the HFOC and PLT detector data leads to a systematic uncertainty of 0.2\%.
In the 2018 analysis, three detectors are available.
The uncertainty due to the combination-to-combination variation is 0.2\%, while the luminometer dependency is 0.1\%.

\begin{figure}[!htp]
\centering
\includegraphics[width=0.47\textwidth]{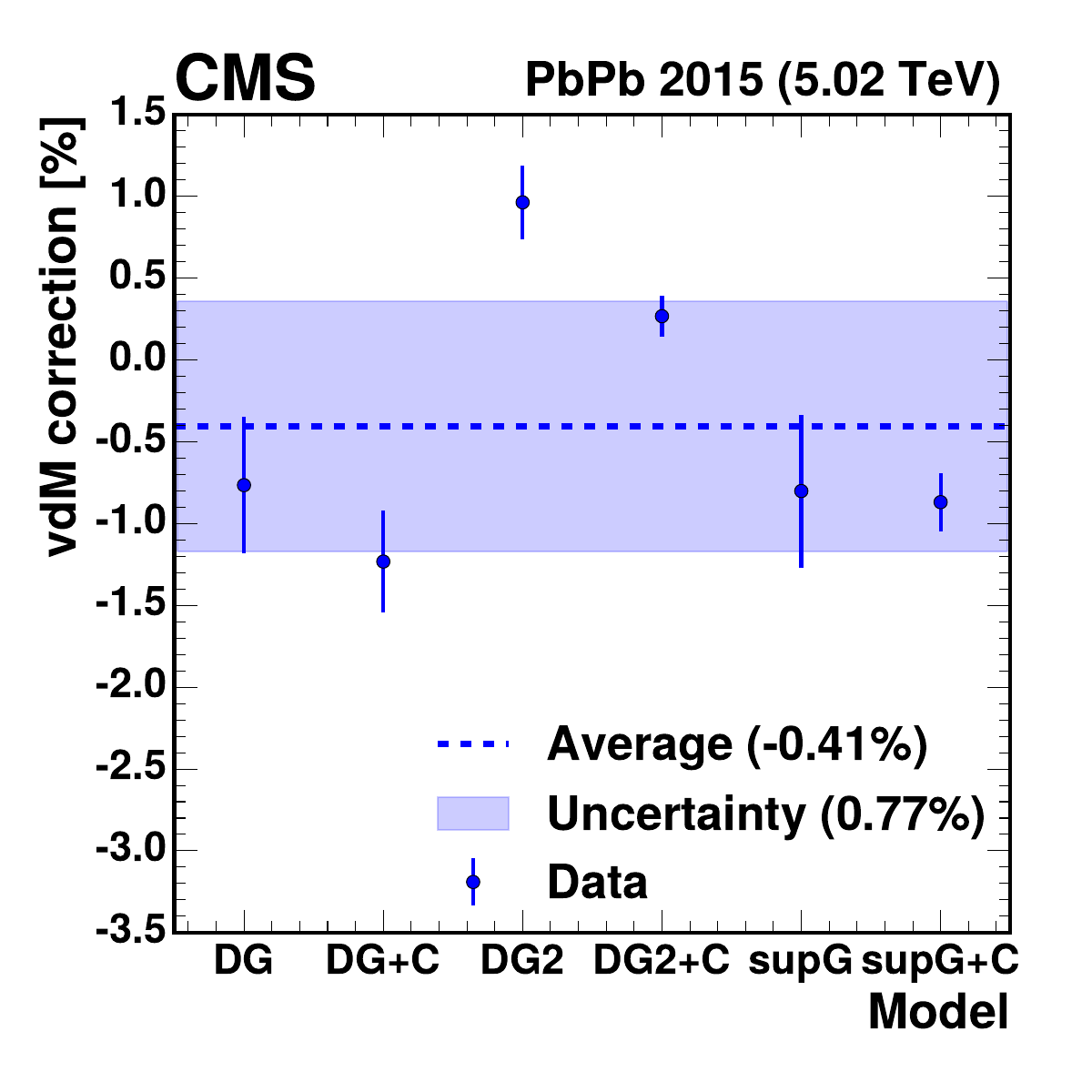}
\caption{%
    Model dependence of the factorizability correction in 2015 for DG, DG2, and supG fit models without and with an extra constant (the latter are labeled as ``+C'').
    The points represent the mean values, with the error bars defined by the RMS over residual OD variations and luminometers of the BCID-averaged corrections.
    The dashed line indicates the average and the shaded band shows the RMS range around the average for the six models.
}
\label{fig:xyCorr:all_corrections_left}
\end{figure}

A significant source of bias comes from the choice of the analytic model for the 2D bunch overlap distribution.
In 2015, a 0.8\% variation around the central value is observed depending on the model choice as illustrated in Fig.~\ref{fig:xyCorr:all_corrections_left}.

\begin{figure}[!htp]
\centering
\includegraphics[width=0.47\textwidth]{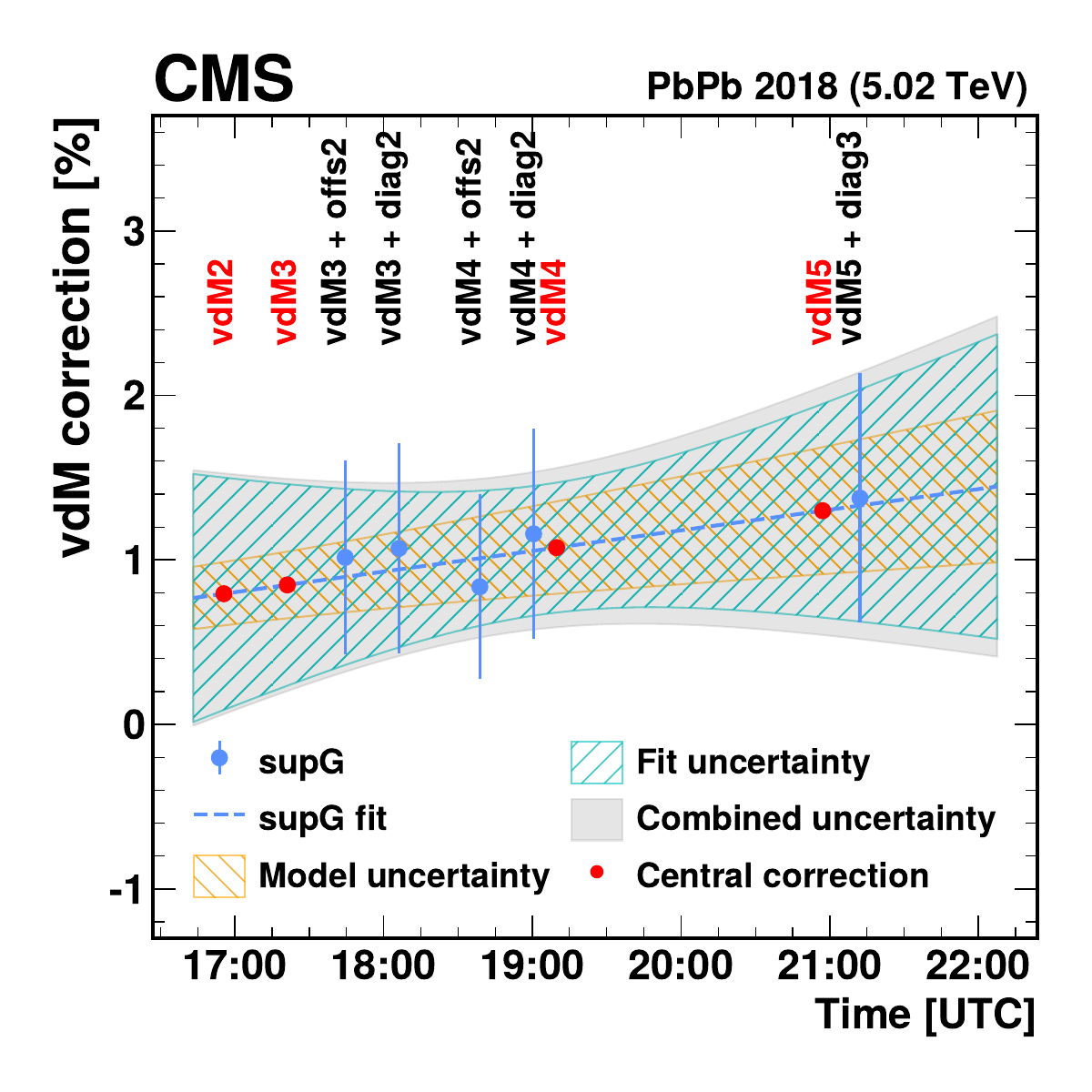}
\caption{%
    Time dependence of the factorizability correction in 2018 for fill 7443, based on the scan combinations vdM3+offset2, vdM3+diag2, offset2+vdM4, diag2+vdM4, and vdM5+diag3 using the supG model as baseline.
    The error bars show the combined uncertainty.
    The dashed blue line shows the trend line over the scan combinations, with the solid gray area denoting the total uncertainty.
    The hatched areas denote contributions to the uncertainty due to the model choice including supG, DG, SG+supG, and supG+supG (orange) and due to the fit (blue) uncertainty of the trend line.
}
\label{fig:xyCorr:all_corrections_right}
\end{figure}

{\tolerance=800
Similarly the model uncertainty is 0.5\% for fill 7442 in 2018, with four models considered (SupG, DG, SG+SupG, and SupG+SupG).
The model choice depends on the year as different functions provided good fits of the data.
The several scan combinations available in fill 7443 allow for a study of the time dependence of the factorizability.
This is done by performing a trend line fit.
Figure~\ref{fig:xyCorr:all_corrections_right} summarizes the estimated correction and total uncertainty.
\par}

The straight line fit takes into account the covariance of the scan combinations emerging due to shared constituent scans. The supG fit gives the most stable results and is chosen for the central values.
The additional fit models (DG, SG+supG, supG+supG) with similar \chisqred contribute to the uncertainty band.

The dominant uncertainty in the factorization bias measurement arises from the closure of the method, \ie, the accuracy with which a bias can be measured.
This is studied using simulated pseudo-experiments with different true distributions for the bunch overlap shape.
The closure is very sensitive to the statistical power of the data.
Due to the very low pileup of the \PbPb runs, the statistical uncertainty in the detector rates at the peak is around 8--10\% for vdM and diagonal scans, and about 17--20\% for offset scans.
The best performing supG model gives a bias of $-0.16\pm0.05\%$ in the 2015 analysis.
The largest nonclosure of $-0.6\pm0.5\%$ is found for the DG model with separate means due to its large number of free parameters.
Conservatively, we take 0.8\% as the nonclosure uncertainty.

The total uncertainty in the factorizability correction is 1.1\% in both years.
In 2015 as the derived bias is significantly smaller than its uncertainty, no correction was applied.
In 2018, a time dependent correction was utilized.

\section{Visible cross section results}
\label{sec:sigvis}

\begin{figure}[!ht]
\centering
\includegraphics[width=0.48\textwidth]{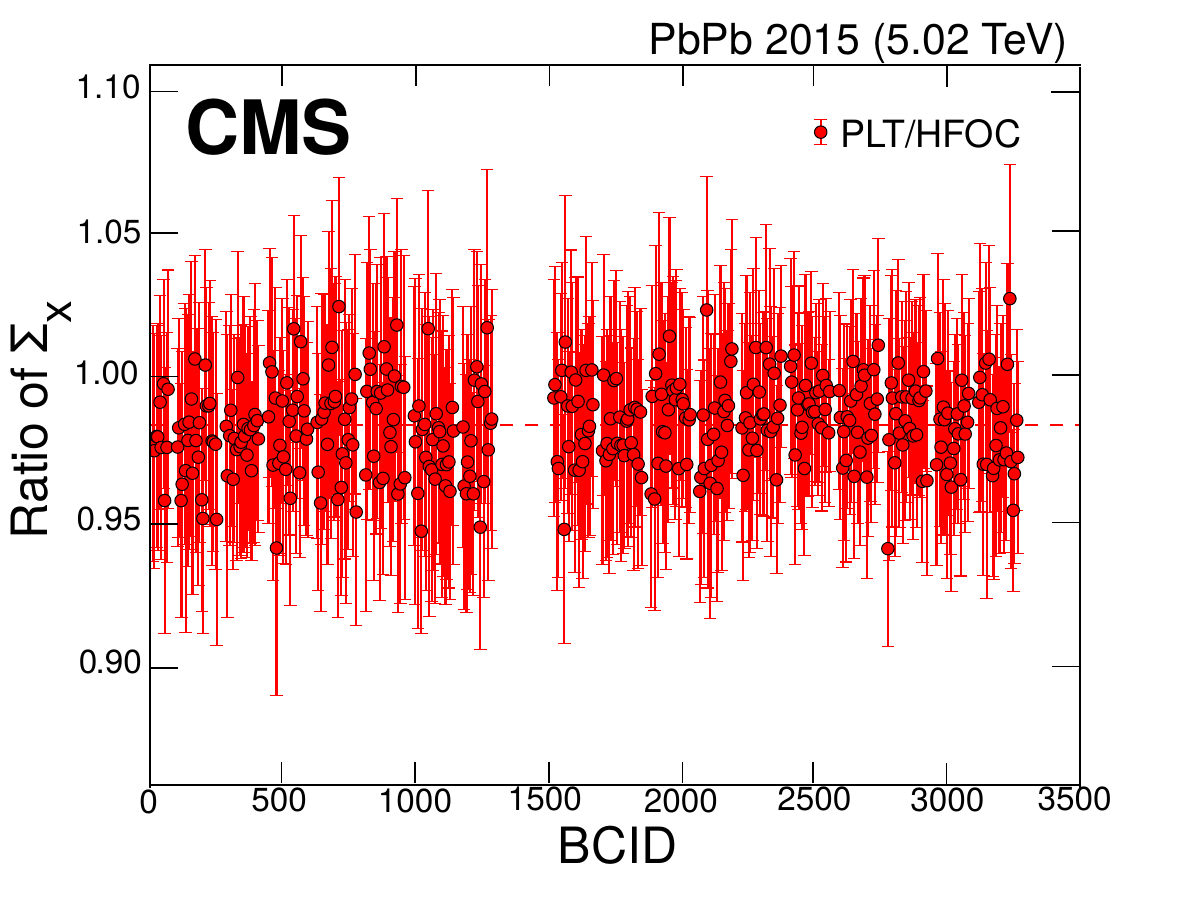}%
\hfill%
\includegraphics[width=0.44\textwidth]{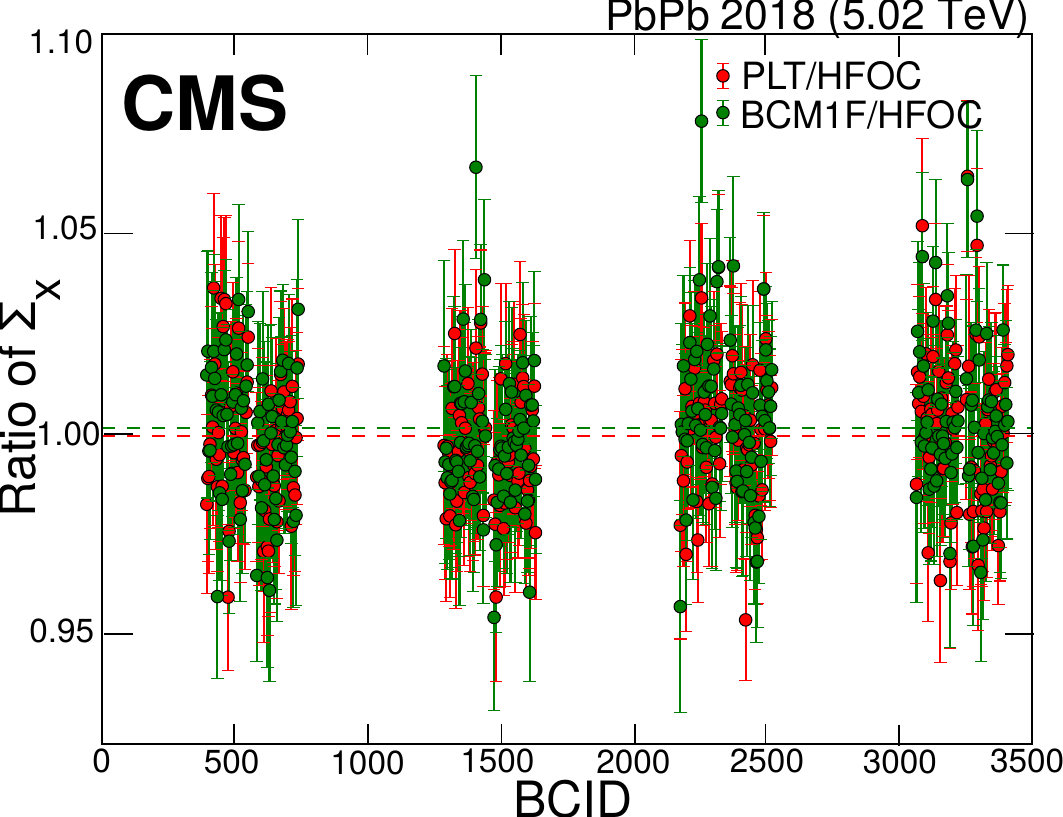}
\caption{%
    Cross-detector comparison of the fitted \Sigx as a function of BCID for the 2015 vdM3 (\cmsLeft) and 2018 vdM5 (\cmsRight) scans.
    The error bars represent the statistical uncertainty.
    The horizontal dashed lines indicate the averaged values of the corresponding set of points.
}
\label{fig:capsigma_ratio_X}
\end{figure}

Examples of cross-detector comparisons of the fitted \Sigx are shown in Fig.~\ref{fig:capsigma_ratio_X} as functions of BCID.
In 2015, we find smaller values with PLT than with HFOC, which might be related to residual contributions from not fully subtracted backgrounds for HFOC.
In 2018, all luminometers measure compatible \Sigx values,
and differences between any two luminometers are statistically consistent with zero.

\begin{figure*}[!ht]
\centering
\includegraphics[width=0.32\textwidth]{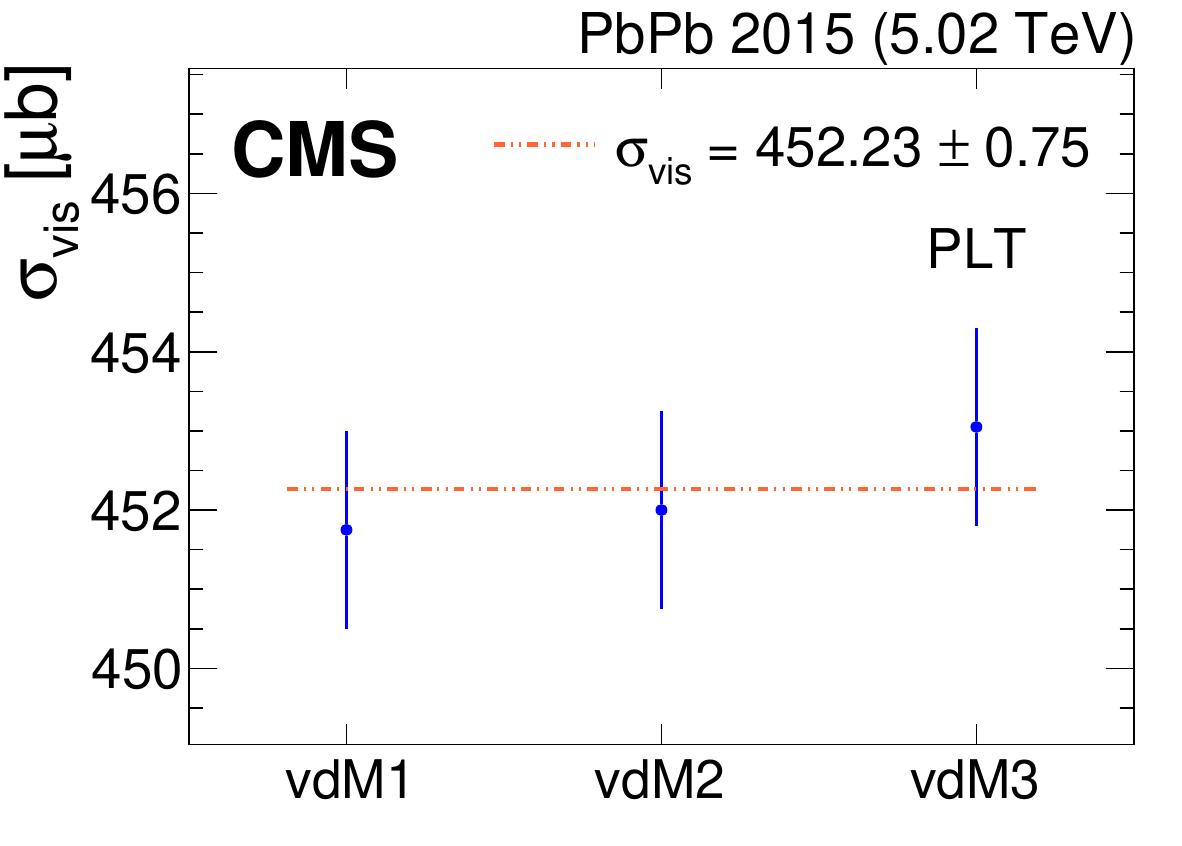}%
\hspace*{0.02\textwidth}%
\includegraphics[width=0.32\textwidth]{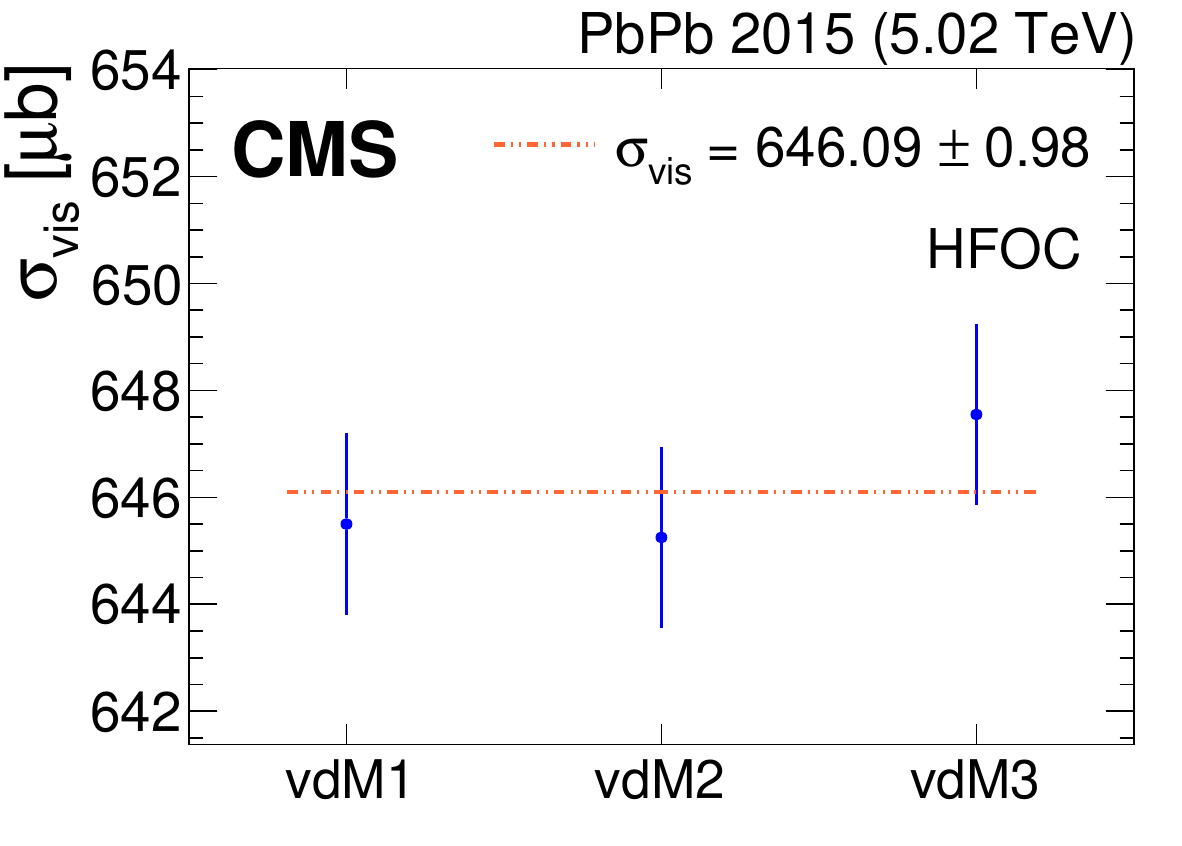}\\
\includegraphics[width=0.32\textwidth]{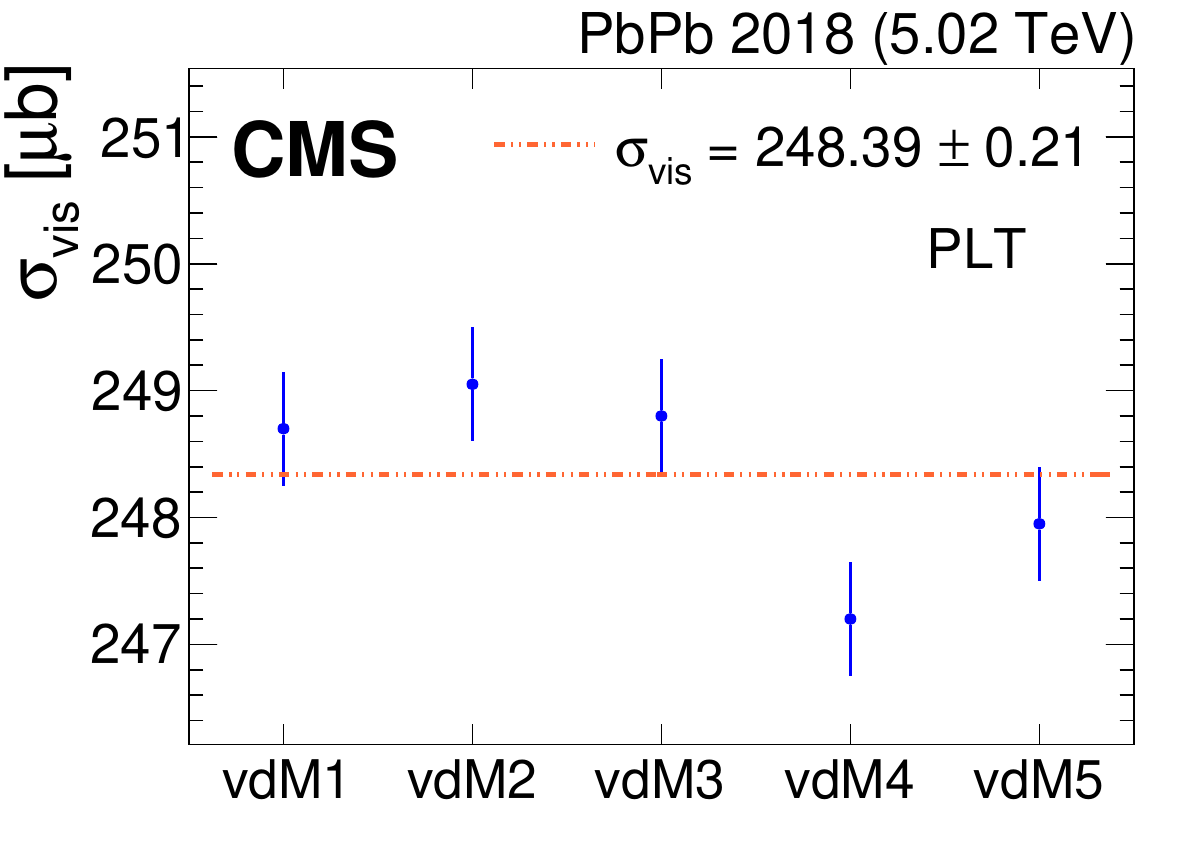}%
\hfill%
\includegraphics[width=0.32\textwidth]{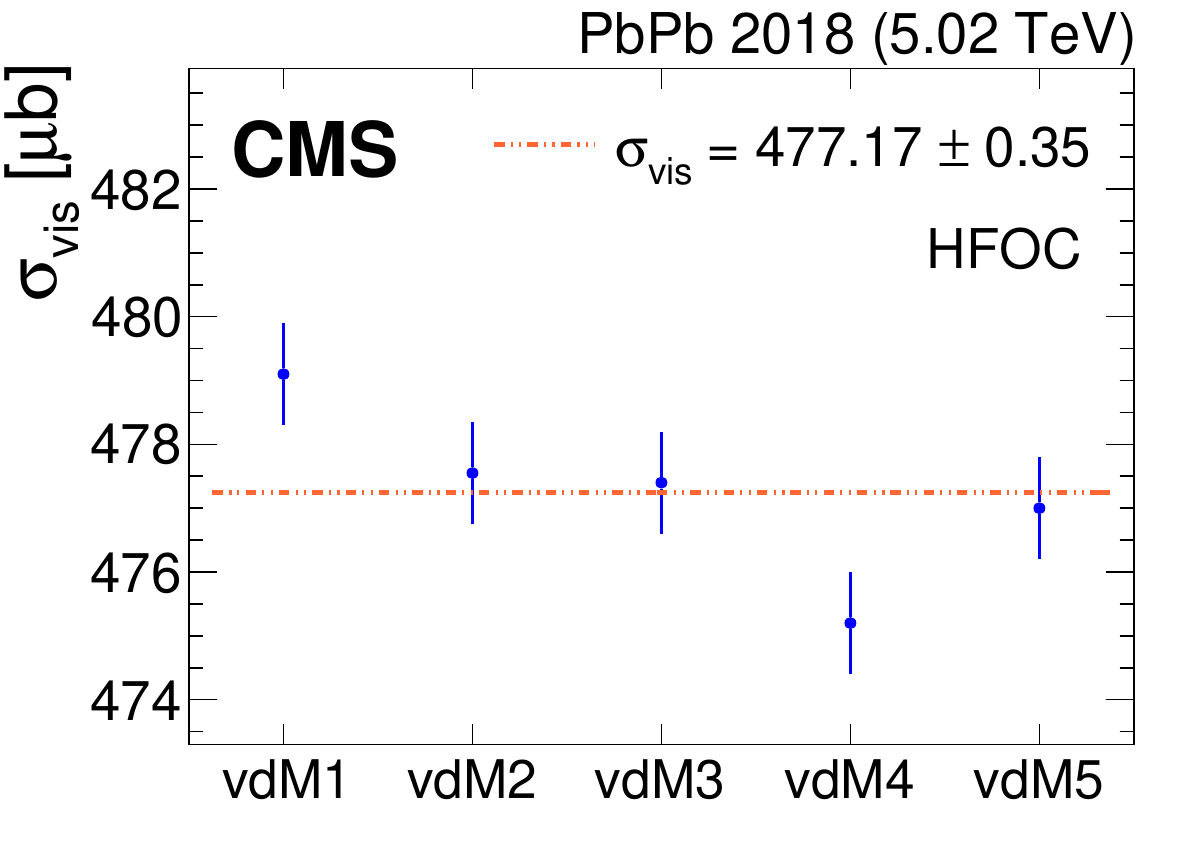}%
\hfill%
\includegraphics[width=0.32\textwidth]{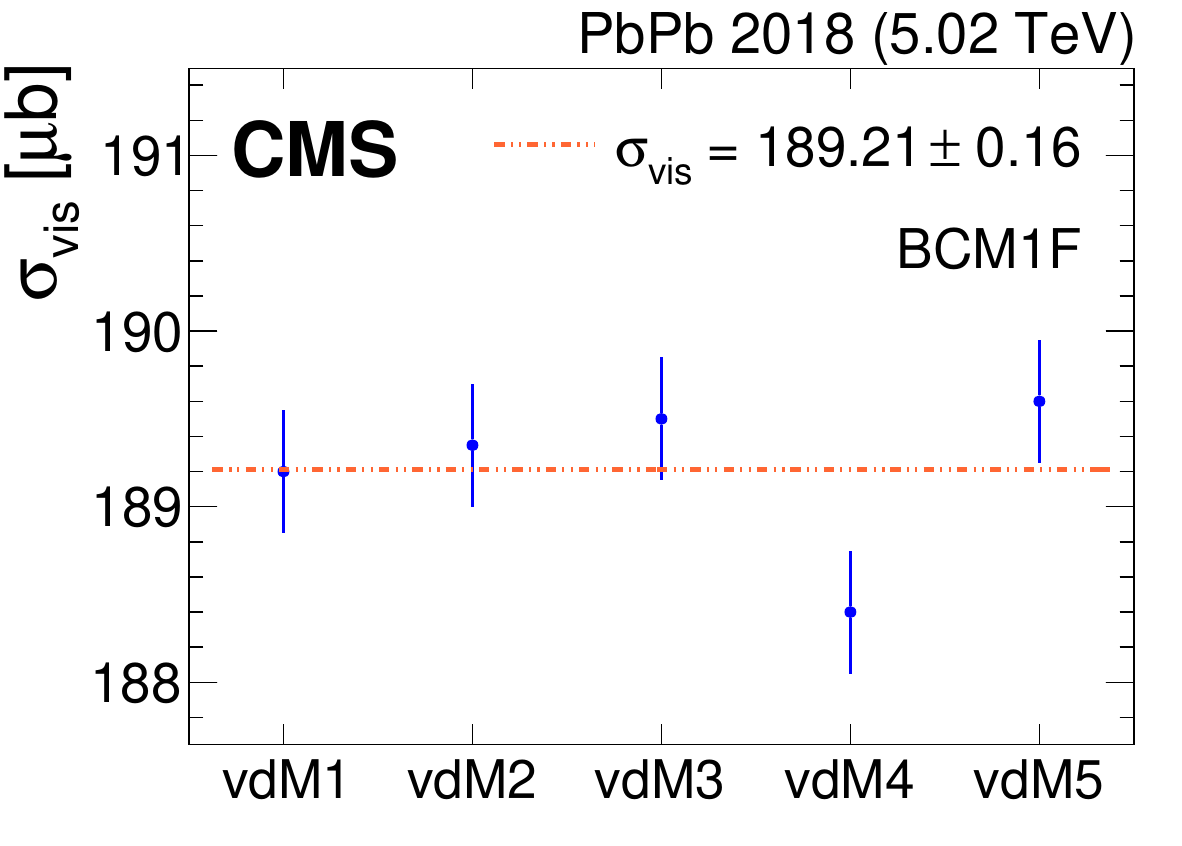}
\caption{%
    The measured \sigmaVis in 2015 (upper row) and 2018 (lower row) calculated by averaging over all BCIDs for each vdM scan pair after all corrections are applied.
    From left to right the results are given for PLT, HFOC, and in 2018 also for BCM1F.
    The error bars represent the standard deviation of the measurements per BCID, accounting for statistical fluctuations.
    To calculate the central value of \sigmaVis, the uncertainty-weighted average of the individual measurements for the various vdM scans are calculated along with its uncertainty.
}
\label{fig:sigma_vis_allscan_perdet_allcorrs}
\end{figure*}

The calculated \sigmaVis values for 2015 and 2018 are given for each vdM scan in Fig.~\ref{fig:sigma_vis_allscan_perdet_allcorrs}.
The uncer\-tainty-weighted average of the measured values for the various vdM scans with its uncertainty is quoted in the plots and taken as the central value for the luminosity calculation.

Good agreement is found among the separate measurements from various 2015 vdM scans.
A dedicated study to assess the scan-to-scan and bunch-to-bunch variations in 2015 shows that the maximal deviation from the average in both cases is covered by the relevant uncertainties (the vdM fit uncertainty for the bunch-to-bunch variation, and the standard deviation over the BCIDs for the scan-to-scan comparisons).
Thus, no additional uncertainty is assigned.

{\tolerance=800
In 2018, no significant variation of \sigmaVis compared with the fit uncertainties is found for the different bunches, and thus the 0.1\% uncertainty in the average value is assigned as the statistical (or bunch-to-bunch variation) uncertainty.
The scan-to-scan variations of the \sigmaVis values in Fig.~\ref{fig:sigma_vis_allscan_perdet_allcorrs} are larger than the uncertainty of the uncertainty-weighted average over the scan pairs (where the input uncertainties are taken as the uncertainty in the average over the BCIDs).
In particular, vdM4 yields consistently lower \sigmaVis values than the other vdM scans.
Thus, the scan-to-scan systematic uncertainty is calculated as the maximum difference between the average and the individual \sigmaVis measurements for the 2018 vdM1--5 scans.
It is found to be about 0.5\%, similar for all luminometers.
\par}

\section{Cross-detector consistency and stability}
\label{sec:stability}

\subsection{Luminosity monitoring}
\label{sec:stability_lumi}

The comparison of the response of independent luminometers throughout physics data taking reveals possible time- and rate-dependent effects on the \sigmaVis calibration.
As PLT is the luminometer least affected by noncollision backgrounds, it is chosen as the primary luminometer for both the 2015 and 2018 \PbPb data-taking periods.

The HFOC subdetector has out-of-time contributions that are not arising from collisions within the window of the bunch crossing.
These contributions are split into ``type 1'' and ``type 2'' effects, representing spillover of electronic signals and real additional response from material activation, respectively.
They are subtracted for all bunch crossings before the total instantaneous luminosity is computed, following a procedure similar to that of Ref.~\cite{CMS:2021xjt}.
The residual rate in empty or unpaired BCIDs is approximately zero illustrating the good performance of the correction.
The type 1 and 2 residuals are generally smaller than 0.2 and 0.5\% of the head-on rate, respectively.

\begin{figure*}[!ht]
\centering
\includegraphics[width=0.48\textwidth]{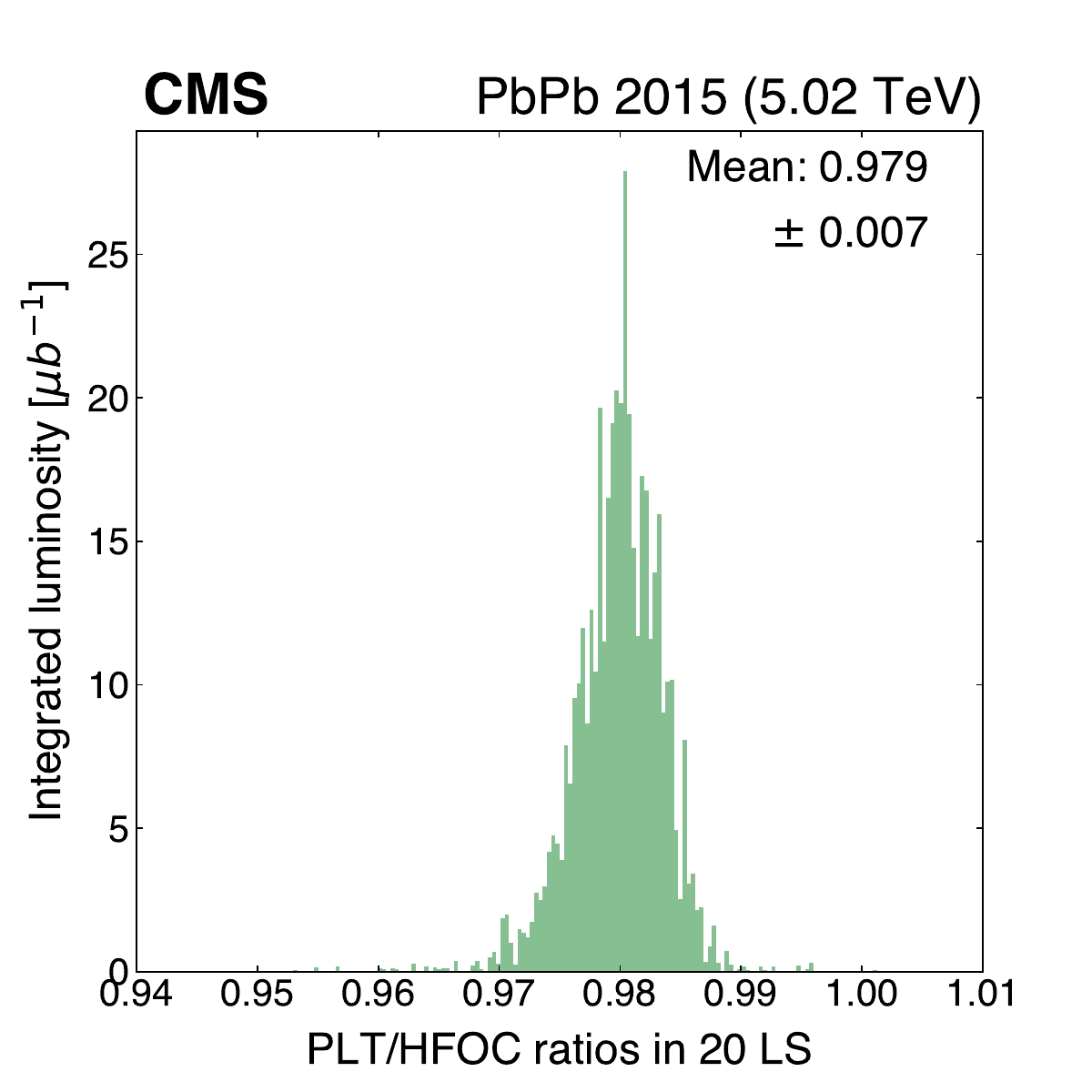}%
\hfill%
\includegraphics[width=0.47\textwidth]{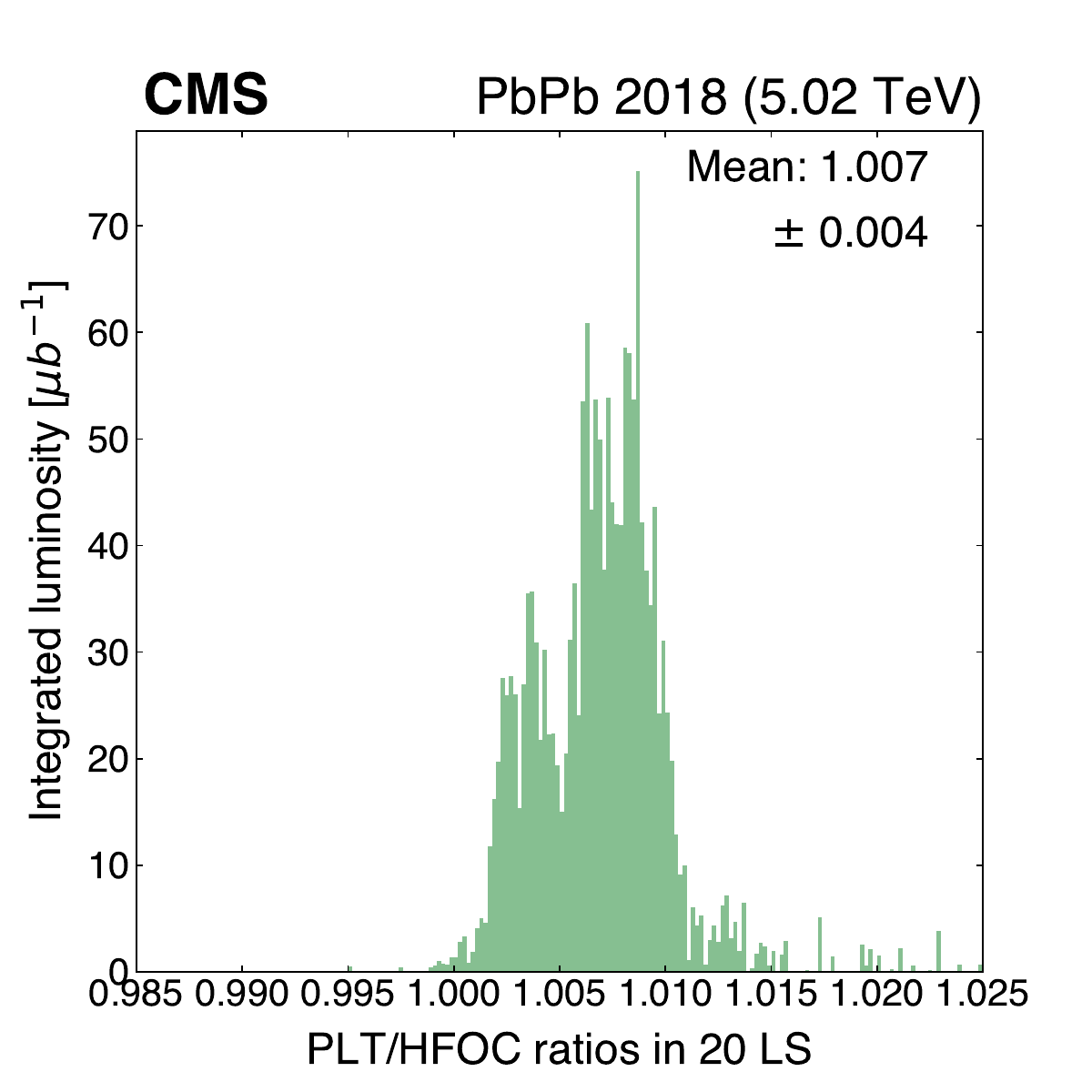} \\[1ex]
\includegraphics[width=\textwidth]{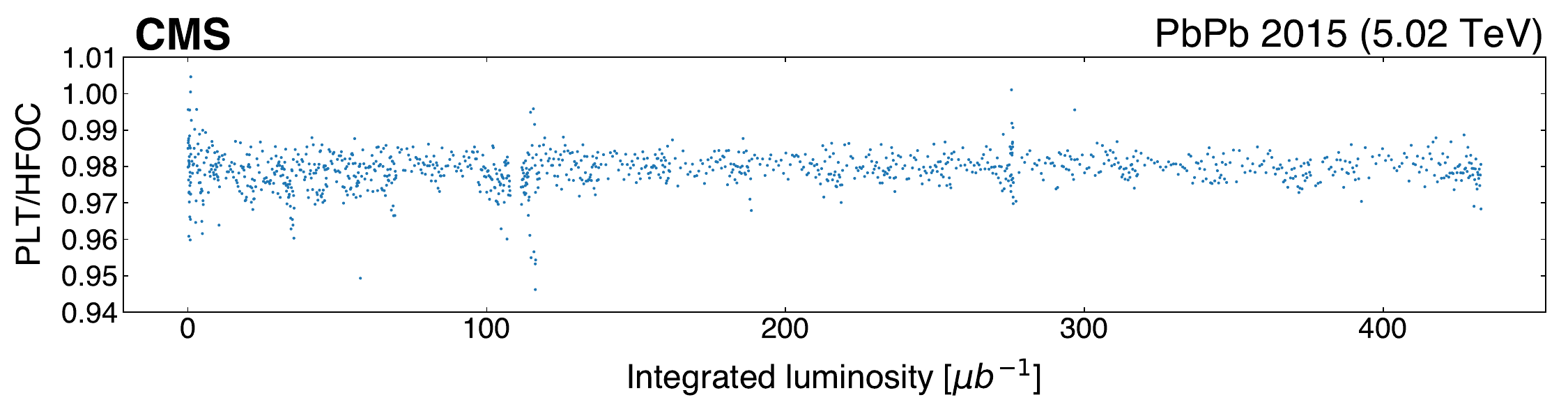} \\[1ex]
\includegraphics[width=\textwidth]{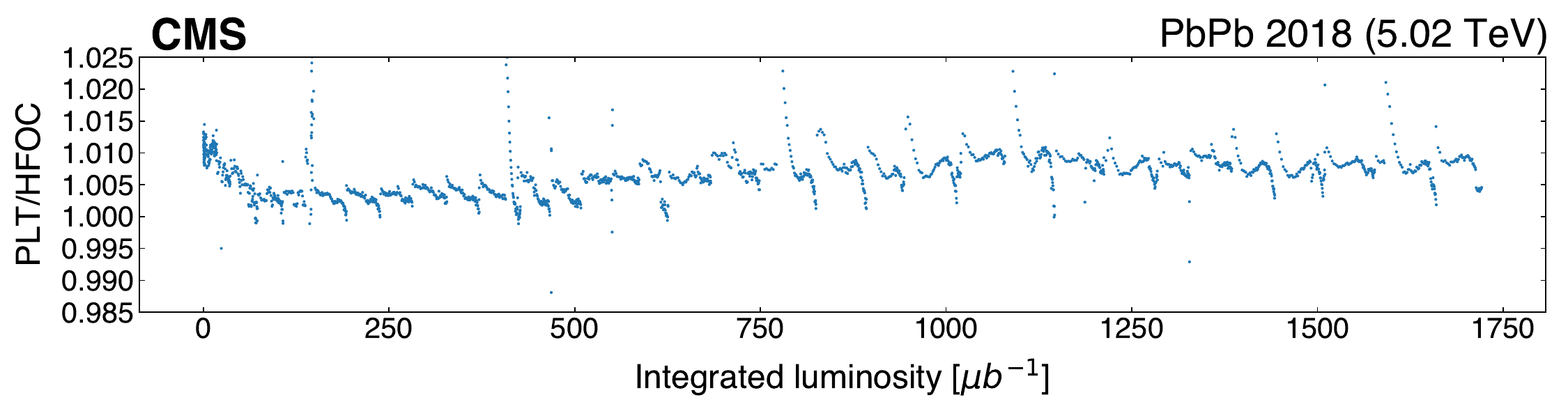}
\caption{%
    Cross-detector comparison of the measured luminosities using the PLT over HFOC luminosity ratio, cumulatively (upper) and as a function of integrated luminosity (middle and lower), for 2015 (upper left and middle) and 2018 (upper right and lower). The luminosity is integrated per 20 LS.
}
\label{fig:PLTOverHFOffline}
\end{figure*}

{\tolerance=800
In 2015, the luminosity ratio between PLT and HFOC during the nonscanning periods of the vdM fill is studied in two time windows.
The larger deviation from unity is found to be about 2.5\%, which is assigned as the systematic uncertainty due to the cross-detector inconsistency.
To estimate the detector stability during the 2015 physics runs, the ratio of the integrated luminosity as measured by HFOC over PLT, as shown in Fig.~\ref{fig:PLTOverHFOffline} (left), is studied.
The mean value of this distribution, representing the difference in the measured luminosity by the two detectors during the 2015 physics runs, shows a 2.1\% deviation from unity, which is already covered by the cross-detector consistency uncertainty.
The standard deviation of the ratio over the full 2015 \PbPb data taking, \ie, 0.7\%, is considered as the systematic uncertainty accounting for the cross-detector stability.
\par}

Similar to the 2015 evaluation, the cross-detector consistency for 2018 is evaluated in nonscanning periods of the vdM fill and found to be 0.4\%, which is assigned as systematic uncertainty.
The ratio of the PLT and HFOC luminosity measurements during the 2018 physics runs is shown in Fig.~\ref{fig:PLTOverHFOffline} (right).
The mean ratio is 0.7\% higher than unity, larger than the 0.4\% difference observed in the vdM fill.
The standard deviation of the ratio is 0.4\%.
The systematic uncertainty related to cross-detector stability is estimated using the quadratic sum of these mean and standard deviation, resulting in a value of 0.8\%, which corresponds to a range that encloses most of the various short- and long-term differences between the PLT and HFOC results.

\subsection{Emittance scans}

\begin{figure}[!ht]
\centering
\includegraphics[width=0.48\textwidth]{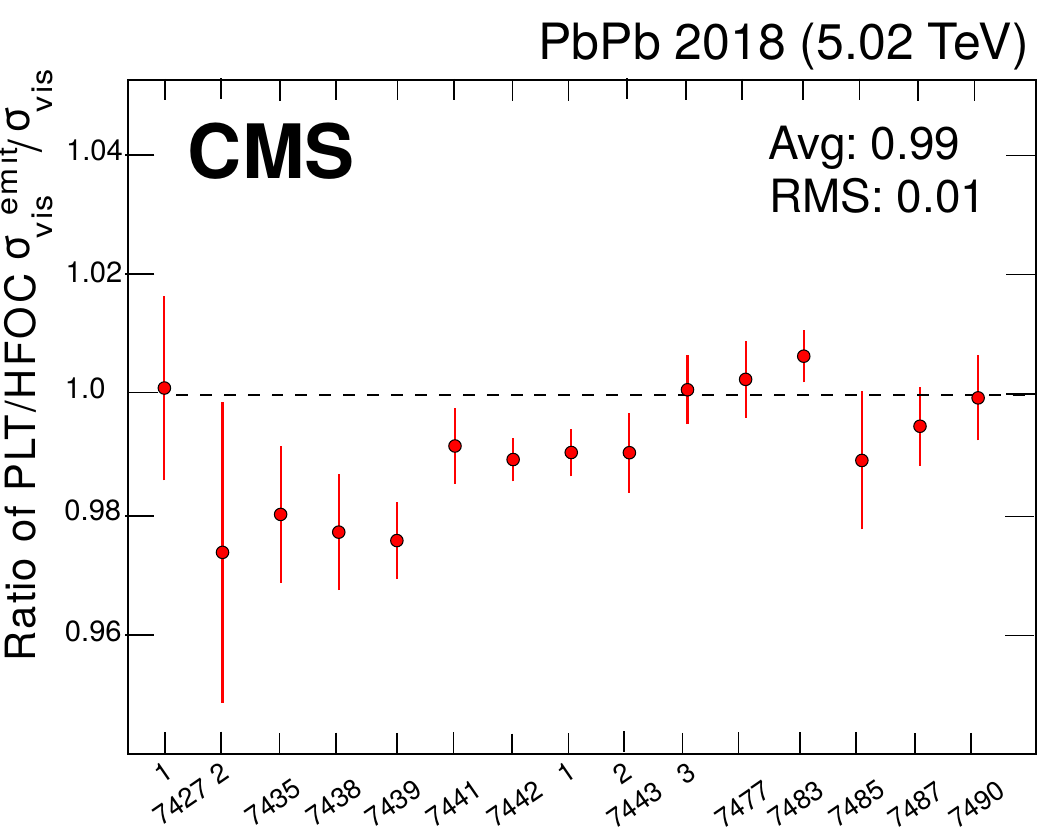}
\caption{%
    The ratio of \sigmaVis-normalized PLT to HFOC cross sections from emittance scans recorded during fills 7442--7443 and routine data-taking conditions in 2018 as indicated with corresponding fill numbers.
    The error bars indicate the statistical uncertainty.
}
\label{fig:FOM}
\end{figure}

The stability is independently investigated using emittance scans performed during regular physics conditions in 2018, similarly to those conducted in fills 7442--7443.
Specifically, they consisted of nine points with 10\unit{s} integration time at each scan point and maximum beam separation
of approximately 4\sigmab.
These scans last only a few minutes and were typically performed at the beginning and at the end of the fill (unless the fill is dumped prematurely).
Because of the smaller beam separation and the reduced amount of data at each scan point, the \sigmaVis values estimated from emittance scan pairs (denoted as $\sigmaVis^{\text{emit}}$) are not as precise as those from vdM scan pairs, but provide a useful relative measurement.
Excellent consistency with the stability evaluation of Section~\ref{sec:stability_lumi} is found, as shown in Fig.~\ref{fig:FOM}, where the PLT over HFOC $\sigmaVis^{\text{emit}}/\sigmaVis$ ratio, corrected for noncollision rate, is presented for emittance scans, along with the mean and the RMS variation.
The beam-condition-related effects cancel out in the double ratio and thus it well quantifies the cross-detector stability.

\section{Total luminosity correction and uncertainty}
\label{sec:syserr}

{\tolerance=800
For each data set, final rate corrections and final calibrations are applied to data in small time windows of $2^{18}$ LHC orbits, so-called ``luminosity sections'' (LS) of approximately 23 seconds.
All measurements are summed to derive a total integrated luminosity measurement.
The contributions to the systematic uncertainty in the integrated luminosity are divided into two general categories:
\begin{itemize}
\item ``normalization'' uncertainty in the absolute luminosity scale, \sigmaVis, determined from the vdM scan procedure;
\item ``integration'' uncertainty associated with \sigmaVis variations over time (stability) and due to out-of-time contributions to the measured rate.
\end{itemize}
The magnitudes of the corrections applied to the absolute normalization from the vdM calibration are listed in Table~\ref{tab:systematics}, and Table~\ref{tab:SystematicError} summarizes all the uncertainties.
The dominant sources of normalization uncertainty are associated with transverse factorizability, and in 2015 the cross-detector consistency in the vdM fill head-on data taking.
The dominant sources of integration uncertainty arise from stability of the primary (typically PLT) relative
to the secondary (typically HFOC) luminosity measurements over the course of each period (as discussed in Section~\ref{sec:stability}).
\par}

{\tolerance=800
Several sources of normalization uncertainty are considered to be correlated for the years studied when the underlying physics or instrumental source of the potential bias is the same, or the scan procedures and analysis methodology are similar between the two vdM calibrations.
The sources of the normalization uncertainty that are not correlated between the two vdM programs, as they are partly statistical or random in nature, are the random (linear) orbit drift, along with the scan-to-scan, and statistical or bunch-to-bunch variations in the measured \sigmaVis, as well as the cross-detector consistency.
The latter three are collectively referred to as ``result consistency'' in Table~\ref{tab:SystematicError}.
The noncollision rate is also considered as uncorrelated because different methods are used to investigate the systematic uncertainty for the two years.
The dominant uncertainty due to transverse factorizability of the bunch particle densities is due to the nonclosure of the method driven by the large statistical uncertainty of the data, and is thus treated as uncorrelated between the years.
\par}

{\tolerance=800
Among the sources of integration uncertainty, the out-of-time contribution corrections are treated identically in the two data sets, and thus this source of systematic uncertainty is correlated.
The stability assessment, on the other hand, is based on cross-detector comparisons.
Although the primary and secondary luminometer is the same in each data set, since the source of instability cannot be assessed and contains time-dependent features, the corresponding uncertainty is considered as not correlated.
\par}

\begin{table}[!ht]
\centering
\topcaption{%
    Summary of the BCID-averaged corrections to \sigmaVis.
    Where applicable, the average correction applied to \sigmaVis of PLT is shown.
}
\label{tab:systematics}
\renewcommand{\arraystretch}{1.1}
\begin{tabular}{lcc}
    \hline
    \multirow{2}{*}{Source} & \multicolumn{2}{c}{Impact on \sigmaVis [\%]} \\[-1pt]
    & 2015 & 2018 \\
    \hline
    Ghost and satellite charge & $+$(8.2--8.6) & $+$(2.5--4.7) \\
    Noncollision rate & $-2.5$ & $-0.6$ \\
    Length scale calibration & $-2.4$ & $-1.5$ \\
    Transverse factorizability & \NA & $+$(0.9--1.3) \\
    Random orbit drift & $+0.5$ & $-0.1$ \\
    Systematic orbit drift & \NA & \NA \\
    Beam-beam effects & \NA & \NA \\
    \hline
\end{tabular}
\end{table}

\begin{table}[!ht]
\centering
\topcaption{%
    Summary of contributions to the relative systematic uncertainty in the integrated luminosity.
    The systematic uncertainty is divided into groups affecting the description of the vdM profile and the bunch population product measurement (normalization), and the measurement of the rate in physics running conditions (integration).
    The last column specifies the correlation of the systematic source among years.
}
\label{tab:SystematicError}
\renewcommand{\arraystretch}{1.1}
\cmsTable{\begin{tabular}{lccc}
    \hline
    Source & 2015 [\%] & 2018 [\%] & Corr. \\
    \hline
    \multicolumn{3}{c}{Normalization uncertainty} \\
    \multicolumn{1}{l}{\textit{Bunch population}}\\
    \hspace{+2mm} Ghost and satellite charge & 0.3 & 0.5 & Yes \\
    \hspace{+2mm} Beam current calibration & 0.2 & 0.2 & Yes \\
    \multicolumn{1}{l}{\textit{Noncolliding bunches}}\\
    \hspace{+2mm} Noncollision rate & 0.5 & 0.2 & No \\
    \multicolumn{1}{l}{\textit{Beam position monitoring}}\\
    \hspace{+2mm} Random orbit drift & 0.5 & 0.1 & No \\
    \hspace{+2mm} Systematic orbit drift & 0.2 & 0.2 & Yes \\
    \multicolumn{1}{l}{\textit{Beam overlap description}}\\
    \hspace{+2mm} Length scale calibration & 0.5 & 0.5 & Yes \\
    \hspace{+2mm} Beam-beam effects & 0.2 & 0.3 & Yes \\
    \hspace{+2mm} Transverse factorizability & 1.1 & 1.1 & No \\
    \multicolumn{1}{l}{\textit{Result consistency}}\\
    \hspace{+2mm} Cross-detector consistency &2.5 &0.4 & No \\
    \hspace{+2mm} Scan-to-scan variation &\NA & 0.5 & No \\
    \hspace{+2mm} Statistical uncertainty & 0.2 & 0.1 & No \\[\cmsTabSkip]
    \multicolumn{3}{c}{Integration uncertainty} \\
    \multicolumn{1}{l}{\textit{Detector performance}}\\
    \hspace{+2mm} Cross-detector stability & 0.7 & 0.8 & No \\
    \multicolumn{1}{l}{\textit{Noncolliding bunches}}\\
    \hspace{+2mm} Noncollision rate & 0.1 & 0.1 & Yes \\[\cmsTabSkip]
    Total normalization uncertainty & 2.9 & 1.5 & \NA \\
    Total integration uncertainty & 0.7 & 0.8 & \NA \\
    Total uncertainty & 3.0 & 1.7 & \NA \\
    \hline
\end{tabular}}
\end{table}

\section{Conclusions}
\label{sec:summary}

{\tolerance=800
The measurement of the luminosity delivered to the CMS experiment during the lead-lead (\PbPb) data-taking periods in 2015 and 2018 at a nucleon-nucleon center-of-mass energy of 5.02\TeV is presented for the first time.
Three subdetectors are used, listed in the order of their priority to provide the luminosity measurement: the pixel luminosity telescope, the forward hadron calorimeter, and the fast beam conditions monitor.
Two groups of uncertainties are considered for effects related to ``normalization'' and ``integration''.
The former concerns the visible cross section as determined from the van der Meer (vdM) scan procedure, and the latter the stability and quality of the measurements by the luminosity subdetectors under the \PbPb data-taking conditions.
The dominant sources of uncertainty contributing to the normalization and integration, as estimated in 2018, are associated to the transverse factorizability of the colliding bunch densities and cross-detector stability, respectively.
In 2015, transverse factorizability, cross-detector consistency, and stability represent the main sources of the uncertainty.
The estimated stability in 2018 is consistent with the analysis of short vdM-like scans performed regularly during the \PbPb data-taking period.
The uncertainty in the normalization is 2.9\% in 2015 and 1.5\% in 2018, and it is about 0.7--0.8\% in the integration
in both years. These are treated as uncorrelated and are summed in quadrature.
\par}

When applying the vdM calibration to the entire periods and requiring that the detectors essential for studying final states with muons took high quality data, the total recorded luminosity is $0.433\pm0.013\nbinv$ in 2015, and $1.700\pm0.029\nbinv$ in 2018.
The relative precision of the integrated luminosity measurement is 3.0\% in 2015 and 1.7\% in 2018, and holds also for other choices of the integration period.
Taking into account the correlations among various systematic sources specified in the last column of Table~\ref{tab:SystematicError}, one can evaluate the combined 2015+2018 luminosity measurement.
The results are summarized in Table~\ref{tab:Run2Lumi}.
The combined uncorrelated contribution is mainly determined by the 2018 uncorrelated precision since it is the significantly larger data set.
The combined data set has an integrated luminosity uncertainty of 1.6\%.

\begin{table*}[!ht]
\centering
\topcaption{%
    Summary of the measured recorded luminosity integrated over time periods when the CMS subsystems necessary for physics studies with muons in the final state were in good operational state.
    Also shown are the correlated and uncorrelated sources of uncertainties, which do not depend on the choice of the integration periods.
    The actual amount of recorded data can depend significantly on the data quality requirements imposed on the various subsystems.
}
\label{tab:Run2Lumi}
\renewcommand{\arraystretch}{1.2}
\begin{tabular}{cccccc}
    \hline
    \multirow{2}{*}{Year} & Recorded & \multicolumn{3}{c}{Uncertainty [\nbinvns]} & Uncertainty [\%] \\[-3pt]
    & luminosity [\nbinvns] & corr. & uncorr. & total & total \\
    \hline
    2015 & 0.433 & 0.003 & 0.013 & 0.013 & 3.0 \\
    2018 & 1.700 & 0.014 & 0.026 & 0.029 & 1.7 \\
    Combined & 2.133 & 0.017 & 0.028 & 0.034 & 1.6 \\
    \hline
\end{tabular}
\end{table*}

\begin{acknowledgments}
    We congratulate our colleagues in the CERN accelerator departments for the excellent performance of the LHC and thank the technical and administrative staffs at CERN and at other CMS institutes for their contributions to the success of the CMS effort. In addition, we gratefully acknowledge the computing centers and personnel of the Worldwide LHC Computing Grid and other centers for delivering so effectively the computing infrastructure essential to our analyses. Finally, we acknowledge the enduring support for the construction and operation of the LHC, the CMS detector, and the supporting computing infrastructure provided by the following funding agencies: SC (Armenia), BMBWF and FWF (Austria); FNRS and FWO (Belgium); CNPq, CAPES, FAPERJ, FAPERGS, and FAPESP (Brazil); MES and BNSF (Bulgaria); CERN; CAS, MoST, and NSFC (China); MINCIENCIAS (Colombia); MSES and CSF (Croatia); RIF (Cyprus); SENESCYT (Ecuador); ERC PRG, RVTT3 and MoER TK202 (Estonia); Academy of Finland, MEC, and HIP (Finland); CEA and CNRS/IN2P3 (France); SRNSF (Georgia); BMBF, DFG, and HGF (Germany); GSRI (Greece); NKFIH (Hungary); DAE and DST (India); IPM (Iran); SFI (Ireland); INFN (Italy); MSIP and NRF (Republic of Korea); MES (Latvia); LMTLT (Lithuania); MOE and UM (Malaysia); BUAP, CINVESTAV, CONACYT, LNS, SEP, and UASLP-FAI (Mexico); MOS (Montenegro); MBIE (New Zealand); PAEC (Pakistan); MES and NSC (Poland); FCT (Portugal); MESTD (Serbia); MICIU/AEI and PCTI (Spain); MOSTR (Sri Lanka); Swiss Funding Agencies (Switzerland); MST (Taipei); MHESI and NSTDA (Thailand); TUBITAK and TENMAK (Turkey); NASU (Ukraine); STFC (United Kingdom); DOE and NSF (USA).
       
    \hyphenation{Rachada-pisek} Individuals have received support from the Marie-Curie program and the European Research Council and Horizon 2020 Grant, contract Nos.\ 675440, 724704, 752730, 758316, 765710, 824093, 101115353, 101002207, and COST Action CA16108 (European Union); the Leventis Foundation; the Alfred P.\ Sloan Foundation; the Alexander von Humboldt Foundation; the Science Committee, project no. 22rl-037 (Armenia); the Fonds pour la Formation \`a la Recherche dans l'Industrie et dans l'Agriculture (FRIA-Belgium); the Beijing Municipal Science \& Technology Commission, No. Z191100007219010 and Fundamental Research Funds for the Central Universities (China); the Ministry of Education, Youth and Sports (MEYS) of the Czech Republic; the Shota Rustaveli National Science Foundation, grant FR-22-985 (Georgia); the Deutsche Forschungsgemeinschaft (DFG), among others, under Germany's Excellence Strategy -- EXC 2121 ``Quantum Universe" -- 390833306, and under project number 400140256 - GRK2497; the Hellenic Foundation for Research and Innovation (HFRI), Project Number 2288 (Greece); the Hungarian Academy of Sciences, the New National Excellence Program - \'UNKP, the NKFIH research grants K 131991, K 133046, K 138136, K 143460, K 143477, K 146913, K 146914, K 147048, 2020-2.2.1-ED-2021-00181, TKP2021-NKTA-64, and 2021-4.1.2-NEMZ\_KI-2024-00036 (Hungary); the Council of Science and Industrial Research, India; ICSC -- National Research Center for High Performance Computing, Big Data and Quantum Computing and FAIR -- Future Artificial Intelligence Research, funded by the NextGenerationEU program (Italy); the Latvian Council of Science; the Ministry of Education and Science, project no. 2022/WK/14, and the National Science Center, contracts Opus 2021/41/B/ST2/01369 and 2021/43/B/ST2/01552 (Poland); the Funda\c{c}\~ao para a Ci\^encia e a Tecnologia, grant CEECIND/01334/2018 (Portugal); the National Priorities Research Program by Qatar National Research Fund; MICIU/AEI/10.13039/501100011033, ERDF/EU, "European Union NextGenerationEU/PRTR", and Programa Severo Ochoa del Principado de Asturias (Spain); the Chulalongkorn Academic into Its 2nd Century Project Advancement Project, and the National Science, Research and Innovation Fund via the Program Management Unit for Human Resources \& Institutional Development, Research and Innovation, grant B39G670016 (Thailand); the Kavli Foundation; the Nvidia Corporation; the SuperMicro Corporation; the Welch Foundation, contract C-1845; and the Weston Havens Foundation (USA).
\end{acknowledgments}

\bibliography{auto_generated}
\cleardoublepage \appendix\section{The CMS Collaboration \label{app:collab}}\begin{sloppypar}\hyphenpenalty=5000\widowpenalty=500\clubpenalty=5000\input{LUM-20-002-public-authorlist.tex}\end{sloppypar}
%%% END EDITABLE REGION %%%
% skeleton_end
\end{document}

%% file: LUM-20-002-public-authorlist.tex
\cmsinstitute{Yerevan Physics Institute, Yerevan, Armenia}
{\tolerance=6000
V.~Chekhovsky, A.~Hayrapetyan, V.~Makarenko\cmsorcid{0000-0002-8406-8605}, A.~Tumasyan\cmsAuthorMark{1}\cmsorcid{0009-0000-0684-6742}
\par}
\cmsinstitute{Institut f\"{u}r Hochenergiephysik, Vienna, Austria}
{\tolerance=6000
W.~Adam\cmsorcid{0000-0001-9099-4341}, J.W.~Andrejkovic, L.~Benato\cmsorcid{0000-0001-5135-7489}, T.~Bergauer\cmsorcid{0000-0002-5786-0293}, S.~Chatterjee\cmsorcid{0000-0003-2660-0349}, K.~Damanakis\cmsorcid{0000-0001-5389-2872}, M.~Dragicevic\cmsorcid{0000-0003-1967-6783}, P.S.~Hussain\cmsorcid{0000-0002-4825-5278}, M.~Jeitler\cmsAuthorMark{2}\cmsorcid{0000-0002-5141-9560}, N.~Krammer\cmsorcid{0000-0002-0548-0985}, A.~Li\cmsorcid{0000-0002-4547-116X}, D.~Liko\cmsorcid{0000-0002-3380-473X}, I.~Mikulec\cmsorcid{0000-0003-0385-2746}, J.~Schieck\cmsAuthorMark{2}\cmsorcid{0000-0002-1058-8093}, R.~Sch\"{o}fbeck\cmsAuthorMark{2}\cmsorcid{0000-0002-2332-8784}, D.~Schwarz\cmsorcid{0000-0002-3821-7331}, M.~Sonawane\cmsorcid{0000-0003-0510-7010}, W.~Waltenberger\cmsorcid{0000-0002-6215-7228}, C.-E.~Wulz\cmsAuthorMark{2}\cmsorcid{0000-0001-9226-5812}
% ...
\par}
\vskip\cmsinstskip
\dag:~Deceased\\
$^{1}$Also at Yerevan State University, Yerevan, Armenia\\
$^{2}$Also at TU Wien, Vienna, Austria\\
$^{3}$Also at Ghent University, Ghent, Belgium\\
$^{4}$Also at Universidade do Estado do Rio de Janeiro, Rio de Janeiro, Brazil\\
$^{5}$Also at FACAMP - Faculdades de Campinas, Sao Paulo, Brazil\\
$^{6}$Also at Universidade Estadual de Campinas, Campinas, Brazil\\
$^{7}$Also at Federal University of Rio Grande do Sul, Porto Alegre, Brazil\\
$^{8}$Also at University of Chinese Academy of Sciences, Beijing, China\\
$^{9}$Also at China Center of Advanced Science and Technology, Beijing, China\\
$^{10}$Also at University of Chinese Academy of Sciences, Beijing, China\\
$^{11}$Also at China Spallation Neutron Source, Guangdong, China\\
$^{12}$Now at Henan Normal University, Xinxiang, China\\
$^{13}$Also at University of Shanghai for Science and Technology, Shanghai, China\\
$^{14}$Now at The University of Iowa, Iowa City, Iowa, USA\\
$^{15}$Also at an institute formerly covered by a cooperation agreement with CERN\\
$^{16}$Also at Zewail City of Science and Technology, Zewail, Egypt\\
$^{17}$Also at British University in Egypt, Cairo, Egypt\\
$^{18}$Now at Ain Shams University, Cairo, Egypt\\
$^{19}$Also at Purdue University, West Lafayette, Indiana, USA\\
$^{20}$Also at Universit\'{e} de Haute Alsace, Mulhouse, France\\
$^{21}$Also at Istinye University, Istanbul, Turkey\\
$^{22}$Also at an international laboratory covered by a cooperation agreement with CERN\\
$^{23}$Also at The University of the State of Amazonas, Manaus, Brazil\\
$^{24}$Also at University of Hamburg, Hamburg, Germany\\
$^{25}$Also at RWTH Aachen University, III. Physikalisches Institut A, Aachen, Germany\\
$^{26}$Also at Bergische University Wuppertal (BUW), Wuppertal, Germany\\
$^{27}$Also at Brandenburg University of Technology, Cottbus, Germany\\
$^{28}$Also at Forschungszentrum J\"{u}lich, Juelich, Germany\\
$^{29}$Also at CERN, European Organization for Nuclear Research, Geneva, Switzerland\\
$^{30}$Also at HUN-REN ATOMKI - Institute of Nuclear Research, Debrecen, Hungary\\
$^{31}$Now at Universitatea Babes-Bolyai - Facultatea de Fizica, Cluj-Napoca, Romania\\
$^{32}$Also at MTA-ELTE Lend\"{u}let CMS Particle and Nuclear Physics Group, E\"{o}tv\"{o}s Lor\'{a}nd University, Budapest, Hungary\\
$^{33}$Also at HUN-REN Wigner Research Centre for Physics, Budapest, Hungary\\
$^{34}$Also at Physics Department, Faculty of Science, Assiut University, Assiut, Egypt\\
$^{35}$Also at Punjab Agricultural University, Ludhiana, India\\
$^{36}$Also at University of Visva-Bharati, Santiniketan, India\\
$^{37}$Also at Indian Institute of Science (IISc), Bangalore, India\\
$^{38}$Also at Amity University Uttar Pradesh, Noida, India\\
$^{39}$Also at IIT Bhubaneswar, Bhubaneswar, India\\
$^{40}$Also at Institute of Physics, Bhubaneswar, India\\
$^{41}$Also at University of Hyderabad, Hyderabad, India\\
$^{42}$Also at Deutsches Elektronen-Synchrotron, Hamburg, Germany\\
$^{43}$Also at Isfahan University of Technology, Isfahan, Iran\\
$^{44}$Also at Sharif University of Technology, Tehran, Iran\\
$^{45}$Also at Department of Physics, University of Science and Technology of Mazandaran, Behshahr, Iran\\
$^{46}$Also at Department of Physics, Faculty of Science, Arak University, ARAK, Iran\\
$^{47}$Also at Helwan University, Cairo, Egypt\\
$^{48}$Also at Italian National Agency for New Technologies, Energy and Sustainable Economic Development, Bologna, Italy\\
$^{49}$Also at Centro Siciliano di Fisica Nucleare e di Struttura Della Materia, Catania, Italy\\
$^{50}$Also at Universit\`{a} degli Studi Guglielmo Marconi, Roma, Italy\\
$^{51}$Also at Scuola Superiore Meridionale, Universit\`{a} di Napoli 'Federico II', Napoli, Italy\\
$^{52}$Also at Fermi National Accelerator Laboratory, Batavia, Illinois, USA\\
$^{53}$Also at Lulea University of Technology, Lulea, Sweden\\
$^{54}$Also at Consiglio Nazionale delle Ricerche - Istituto Officina dei Materiali, Perugia, Italy\\
$^{55}$Also at Institut de Physique des 2 Infinis de Lyon (IP2I ), Villeurbanne, France\\
$^{56}$Also at Department of Applied Physics, Faculty of Science and Technology, Universiti Kebangsaan Malaysia, Bangi, Malaysia\\
$^{57}$Also at Consejo Nacional de Ciencia y Tecnolog\'{i}a, Mexico City, Mexico\\
$^{58}$Also at Trincomalee Campus, Eastern University, Sri Lanka, Nilaveli, Sri Lanka\\
$^{59}$Also at Saegis Campus, Nugegoda, Sri Lanka\\
$^{60}$Also at National and Kapodistrian University of Athens, Athens, Greece\\
$^{61}$Also at Ecole Polytechnique F\'{e}d\'{e}rale Lausanne, Lausanne, Switzerland\\
$^{62}$Also at Universit\"{a}t Z\"{u}rich, Zurich, Switzerland\\
$^{63}$Also at Stefan Meyer Institute for Subatomic Physics, Vienna, Austria\\
$^{64}$Also at Laboratoire d'Annecy-le-Vieux de Physique des Particules, IN2P3-CNRS, Annecy-le-Vieux, France\\
$^{65}$Also at Near East University, Research Center of Experimental Health Science, Mersin, Turkey\\
$^{66}$Also at Konya Technical University, Konya, Turkey\\
$^{67}$Also at Izmir Bakircay University, Izmir, Turkey\\
$^{68}$Also at Adiyaman University, Adiyaman, Turkey\\
$^{69}$Also at Bozok Universitetesi Rekt\"{o}rl\"{u}g\"{u}, Yozgat, Turkey\\
$^{70}$Also at Marmara University, Istanbul, Turkey\\
$^{71}$Also at Milli Savunma University, Istanbul, Turkey\\
$^{72}$Also at Kafkas University, Kars, Turkey\\
$^{73}$Now at Istanbul Okan University, Istanbul, Turkey\\
$^{74}$Also at Hacettepe University, Ankara, Turkey\\
$^{75}$Also at Erzincan Binali Yildirim University, Erzincan, Turkey\\
$^{76}$Also at Istanbul University -  Cerrahpasa, Faculty of Engineering, Istanbul, Turkey\\
$^{77}$Also at Yildiz Technical University, Istanbul, Turkey\\
$^{78}$Also at School of Physics and Astronomy, University of Southampton, Southampton, United Kingdom\\
$^{79}$Also at IPPP Durham University, Durham, United Kingdom\\
$^{80}$Also at Monash University, Faculty of Science, Clayton, Australia\\
$^{81}$Also at Universit\`{a} di Torino, Torino, Italy\\
$^{82}$Also at Bethel University, St. Paul, Minnesota, USA\\
$^{83}$Also at Karamano\u {g}lu Mehmetbey University, Karaman, Turkey\\
$^{84}$Also at California Institute of Technology, Pasadena, California, USA\\
$^{85}$Also at United States Naval Academy, Annapolis, Maryland, USA\\
$^{86}$Also at Bingol University, Bingol, Turkey\\
$^{87}$Also at Georgian Technical University, Tbilisi, Georgia\\
$^{88}$Also at Sinop University, Sinop, Turkey\\
$^{89}$Also at Erciyes University, Kayseri, Turkey\\
$^{90}$Also at Horia Hulubei National Institute of Physics and Nuclear Engineering (IFIN-HH), Bucharest, Romania\\
$^{91}$Now at another institute formerly covered by a cooperation agreement with CERN\\
$^{92}$Also at Texas A\&M University at Qatar, Doha, Qatar\\
$^{93}$Also at Kyungpook National University, Daegu, Korea\\
$^{94}$Also at Yerevan Physics Institute, Yerevan, Armenia\\
$^{95}$Also at another international laboratory covered by a cooperation agreement with CERN\\
$^{96}$Also at Imperial College, London, United Kingdom\\
$^{97}$Also at Institute of Nuclear Physics of the Uzbekistan Academy of Sciences, Tashkent, Uzbekistan\\
$^{98}$Also at another institute formerly covered by a cooperation agreement with CERN\\